\documentclass[aps,prd,groupedaddress,nofootinbib]{revtex4}

\def\ga{\mathrel{\raise.3ex\hbox{$>$\kern-.75em\lower1ex\hbox{$\sim$}}}}
\def\la{\mathrel{\raise.3ex\hbox{$<$\kern-.75em\lower1ex\hbox{$\sim$}}}}

\def\lsim{\mathrel{\rlap{\lower4pt\hbox{\hskip1pt$\sim$}}
    \raise1pt\hbox{$<$}}}                
\def\gsim{\mathrel{\rlap{\lower4pt\hbox{\hskip1pt$\sim$}}
    \raise1pt\hbox{$>$}}}                

\usepackage{amsmath}
\usepackage{amssymb}
\usepackage{bbm}
\usepackage{epsfig}
\usepackage{slashed}
\begin{document}
\begin{flushright}
KANAZAWA-11-07\\
SHEP-11-06\\
UT-HET 052
\end{flushright}
\title{Light Charged Higgs bosons at the LHC in 2HDMs}

\author{Mayumi Aoki$^{1}$~\footnote{mayumi@hep.s.kanazawa-u.ac.jp},  Renato Guedes$^{2}$~\footnote{renato@cii.fc.ul.pt}, Shinya Kanemura$^{3}$~\footnote{kanemu@sci.u-toyama.ac.jp}
Stefano Moretti$^{4}$~\footnote{stefano@phys.soton.ac.uk},  Rui Santos$^{2,5}$~\footnote{rsantos@cii.fc.ul.pt} and Kei Yagyu$^{3}$~\footnote{keiyagyu@jodo.sci.u-toyama.ac.jp}}
\affiliation{$^1$Institute for Theoretical Physics, Kanazawa University, Kanazawa 920-1192, Japan.}
\affiliation{$^2$Centro de F\'\i sica Te\' orica e Computacional, Faculdade de Ci\^encias, Universidade de Lisboa, Av. Prof. Gama Pinto 2, 1649-003 Lisboa, Portugal.}
\affiliation{$^3$Department of Physics, The University of Toyama, 3190 Gofuku, Toyama 930-8555, Japan.}
\affiliation{$^4$NExT Institute and School of Physics and Astronomy, University of Southampton, Highfield, Southampton SO17 1BJ, UK.}
\affiliation{$^5$Instituto Superior de Engenharia de Lisboa, Rua Conselheiro Em\'\i dio Navarro 1, 1959-007 Lisboa, Portugal}

\date{\today}

\begin{abstract}
We present a discussion of light charged Higgs boson searches at the Large Hadron Collider (LHC) 
in CP-conserving 2-Higgs Doublet Models (2HDMs). Taking into account all available experimental and 
theoretical constraints we review all possible processes that would allow for a detection of such a 
particle with a mass below the top quark mass. Two different types of processes are analysed: 
one that depends only on $\tan \beta$ and on the charged Higgs boson mass 
because it involves only the charged Higgs boson Yukawa couplings; 
the other that depends on almost all model parameters, mainly due to the presence of Higgs self-couplings. 
We discuss the regions of parameter space of 2HDMs that can be covered by each type of process and 
define some guidelines for experimental searches at the LHC. 
\end{abstract}

\maketitle

\section{Introduction}

\noindent
The Large Hadron Collider (LHC) is operating at a Centre-of-Mass (CM) energy of 7 $TeV$. Soon it will start 
operating at an energy of 14 $TeV$. One of the most striking evidences of physics Beyond the Standard Model 
(BSM) would be the appearance of a charged Higgs boson, $H^\pm$. As light charged Higgs particles,
for which $m_{H^\pm}<m_t$ (the top mass), are simpler to detect, it is timely to perform a study on 
the possible models where these states are still allowed.  The LEP experiments have set a lower limit on 
the mass of a charged Higgs boson, of 79.3$~GeV$ at 95\% Confidence Level (CL), assuming that 
BR$(H^+ \to \tau^+ \nu) + BR(H^+ \to c \bar s)=1$ holds for the possible charged Higgs boson
Branching Ratios (BRs)~\cite{LEP}. This limit becomes stronger 
if BR$(H^+ \to \tau^+ \nu)  \approx 1$ (see \cite{Logan:2009uf} for a discussion). Searches at 
the Tevatron~\cite{Tev} based on $t \bar t$ production with the top decaying via $t \to b H^+$ and assuming 
BR$(H^+ \to \tau^+ \nu) \approx 1$ have yielded a limit of BR$(t \to b H^+) < 0.2$ for a charged Higgs mass 
of 100 $GeV$. Indeed, all models discussed in this work have a BR$(t \to b H^+)$ below the Tevatron limit when 
BR$(H^+ \to \tau^+ \nu)  \approx$ 1\footnote{This conclusion already takes into account other 
experimental constraints.}.  In this study we will concentrate on  the 14 $TeV$ CM energy because, as it
will become clear later on, not only the cross sections grow with energy but also a significant 
luminosity is needed to start probing these models. We therefore defer the study of the 7 $TeV$ case
to a separate publication.    

\noindent
The simplest extensions of the SM that give rise to charged Higgs bosons amount to the addition of an 
extra $SU(2)$ Higgs doublet to the SM field content. The most common CP-conserving 2HDM  has a softly broken 
$Z_2$ symmetry. When this symmetry is extended to the fermions to avoid Flavour Changing Neutral Currents (FCNCs) 
we end up with four~\cite{barger} different models, to be described in detail later, which we will call Type I, 
Type II, Type Y~\cite{KY} and Type X~\cite{KY}  (named I, II, III and IV in~\cite{barger}, respectively). Constraints 
from $B$-physics, and particularly those coming from $b \to s \gamma$~\cite{bsgamma}, have excluded a charged 
Higgs boson with a mass below approximately 300 $ GeV$almost independently of $\tan \beta = v_2/v_1$ -- the ratio of the 
Vacuum Expectation Values (VEVs) of the two doublets - in models Type II and Type Y. Charged Higgs bosons 
with masses as low as 
100 $ GeV$ are instead still allowed in models Type I and X~\cite{Logan:2009uf,KY,Su:2009fz}. The models as 
well as their experimental and theoretical  constraints will be discussed in detail in the next section.  

\noindent
It is important to note that the phenomenology of a light charged Higgs boson, to be discussed in this work 
for charged Higgs searches in specific 2HDM models, is much more general. There is in fact a number of models 
that share a common charged Higgs boson phenomenology for vast regions of the parameter space of those
discussed here.  All such models have in common the fact they have a specific type of 2HDM as submodel for 
Electro-Weak Symmetry Breaking (EWSB). Recently, a number of these scenarios have been discussed in the 
literature~\cite{Aoki:2008av}, wherein the charged Higgs boson BR into leptons is enhanced 
relative to the SM case (Type X). These models provide Dark Matter (DM) candidates naturally and can accommodate 
neutrino oscillations and the strong first order phase transition required for successful baryogenesis while being in 
agreement with all experimental data.

\noindent
The plan of the paper is as follows. The next section is devoted to describe the 2HDM versions we are using and 
their Yukawa Types together with the necessary conditions for the existence of a light charged Higgs boson. 
Section III describes the main production and decay modes of such a light Higgs state at the LHC. Each 
subsection of section III is devoted to the analysis of a specific production process. In Section IV we discuss 
the benchmarks for the search of a light charged Higgs boson at the LHC and draw our conclusions. Finally, one 
Appendix has the detailed description of the parton level analysis of charged Higgs bosons coming from single top 
production processes; a second Appendix details our use of results on charged Higgs boson pair production 
in left-right symmetric models for the purpose of studying the 2HDMs considered here.

\section{The flavour conserving 2HDM}
\label{Sec:2HDMs}
\noindent
We start with a review of the basic 2HDM used in this work. The 2HDM potential chosen here is the most general,
renormalisable and invariant under $SU(2) \otimes U(1)$ that one can build with two complex $SU(2)$ Higgs 
doublets with a softly broken $Z_2$ symmetry. It can be written as
\begin{eqnarray}
 V_{\rm 2HDM}&=& m_1^2 |\Phi_1|^2+m_2^2 |\Phi_2|^2-(m_3^2
  \Phi_1^\dagger \Phi_2 + {\rm h.c.})\nonumber\\
&&  + \frac{1}{2} \lambda_1 |\Phi_1|^4
  + \frac{1}{2} \lambda_2 |\Phi_2|^4
 + \lambda_3 |\Phi_1|^2|\Phi_2|^2
  + \lambda_4 |\Phi_1^\dagger \Phi_2|^2
  + \frac{\lambda_5}{2} \left\{(\Phi_1^\dagger \Phi_2)^2 + {\rm h.c.}
                        \right\},
\end{eqnarray}
where $m_{i}^2$, $i=1,2,3$ and $\lambda_i$ are real (we assume CP conservation). The Higgs doublet fields' VEVs are $v_1$ and $v_2$ and 
satisfy $v_1^2+v_2^2 = v^2 \simeq (246~{{GeV}})^2$. As CP is conserved, and once 
the $SU(2)$ symmetry is broken,  we end up with two CP-even Higgs states, $h$ and $H$, one CP-odd state, $A$, 
two charged Higgs bosons, $H^{\pm}$ and three Goldstone bosons, the latter three
giving mass to the $W^\pm$ and $Z$ gauge bosons. This potential has in principle 8 independent parameters. 
However, because  $v$ is fixed by, e.g., the $W^\pm$ boson mass, only 7 independent parameters remain to be 
chosen, which we take to be  $m_{h}$, $m_{H}$, $m_{A}$, $m_{H^\pm}$, $\tan\beta=v_2/v_1$, $\alpha$ and $M^2$. 
The angle $\beta$ is the rotation angle from the group eigenstates to the mass eigenstates in the CP-odd and 
charged Higgs sector. The angle $\alpha$ is the corresponding rotation angle for the CP-even Higgs sector. 
The parameter $M^2$ is defined as $M^2=m_3^2/(\sin \beta \cos \beta)$ and is a measure of how softly the discrete 
symmetry is broken. The definition of $\alpha$ and the relation among physical scalar masses and coupling 
constants are shown in Ref.~\cite{KOSY} for definiteness.

\noindent
The discrete symmetry imposed on the potential, when extended to the Yukawa Lagrangian, guarantees that 
FCNCs are not present at tree level, as fermions of a given electric charge couple to no more than one Higgs 
doublet~\cite{Glashow}. There are a total of four possible combinations~\cite{barger} 
and therefore four variations of the basic model. Writing the most general Yukawa interaction under the 
$Z_2$ symmetry as
\begin{align}
{\mathcal L}_\text{Yukawa}^\text{2HDM} =
&-{\overline Q}_LY_u\widetilde{\Phi}_uu_R^{}
-{\overline Q}_LY_d\Phi_dd_R^{}
-{\overline L}_LY_\ell\Phi_\ell \ell_R^{}+\text{h.c.},
\end{align}
where $\Phi_f$ ($f=u,d$ or $\ell$) is either $\Phi_1$ or $\Phi_2$, the four independent $Z_2$ charge assignments 
on quarks and charged leptons can be summarised in table~\ref{Tab:type}~\cite{barger,grossmann,Akeroyd:1994ga}.
\begin{table}[h!]
\begin{center}
\begin{tabular}{|c||c|c|c|c|c|c|}
\hline & $\Phi_1$ & $\Phi_2$ & $u_R^{}$ & $d_R^{}$ & $\ell_R^{}$ &
 $Q_L$, $L_L$ \\  \hline
Type I  & $+$ & $-$ & $-$ & $-$ & $-$ & $+$ \\
Type II & $+$ & $-$ & $-$ & $+$ & $+$ & $+$ \\
Type X  & $+$ & $-$ & $-$ & $-$ & $+$ & $+$ \\
Type Y  & $+$ & $-$ & $-$ & $+$ & $-$ & $+$ \\
\hline
\end{tabular}
\end{center}
\caption{Variation in charge assignments of the $Z_2$ symmetry defining the 2HDM Types discussed in this work.} \label{Tab:type}
\end{table}
We define as Type I the model where only the doublet $\Phi_2$ couples to all fermions; 
Type II, the one similar to the  Minimal Supersymmetric Standard Model
(MSSM), is the model where $\Phi_2$ couples to up-type quarks and $\Phi_1$ couples to down-type quarks and 
leptons; a Type Y~\cite{KY} (or Type III) model is built such that $\Phi_2$ couples to up-type quarks and 
to leptons and $\Phi_1$ couples to down-type quarks and finally in a Type X~\cite{KY} (or Type IV model)
$\Phi_2$ couples to all quarks and $\Phi_1$ couples to all leptons.

\noindent
We will now discuss the constraints on these 2HDM Types and in particular the existence of a light charged Higgs 
boson in this context. The parameter space of the aforementioned 2HDMs (or 2HDM Types),
as well as of the family of models with similar 
phenomenology, is limited by experimental data and theoretical constraints (such as vacuum stability and 
perturbative unitarity). It is well known that in the 2HDM
Type II a charged Higgs with a mass below $\approx 300 \, GeV$ is disallowed by the constraints from the 
measurement of $b \to s \gamma$~\cite{bsgamma}. The same bound applies to model Type Y. However, because 
in models Type I and X the same doublet couples to both up- and down-type quarks, the bound can be relaxed 
to 100 $GeV$ or even less depending on the value of $\tan \beta$. In this work we are interested in the light charged Higgs boson mass region, that is, the 
one below the $m_t + m_b$ threshold. Hence, only models Type I and X from a CP-conserving 2HDM 
(and models that contain them as submodels~\cite{Aoki:2008av}) allow for the existence of a light charged 
Higgs state. Finally, we note that we are not considering models that due to a larger particle content, like for instance the MSSM, have the bound on the charged Higgs boson mass relaxed  
in definite regions of the parameter space as a consequence of extra contributions to processes like $b \to s \gamma$. 

\begin{figure}[h!]
\centering
\includegraphics[height=3.6in]{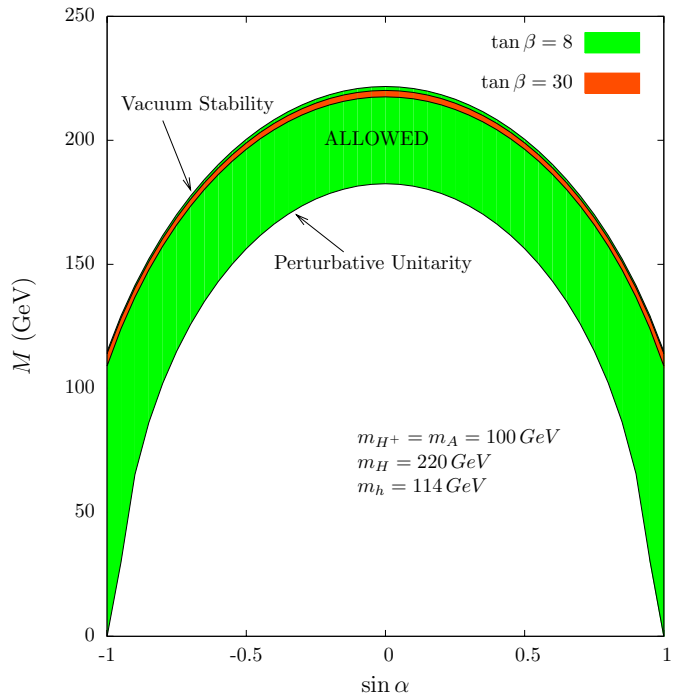}
\caption {Region of parameter space allowed in the plane ($M^2$, $\sin \alpha$)  
when vacuum stability and perturbative unitarity constraints are considered. This plot is
valid for all Yukawa versions of the basic 2HDM prior to enforcing the experimental constraints. }
\label{fig:thconst}
\end{figure}
\vskip -0.1cm
%
\noindent
There are other bounds that constrain the 2HDMs
%
which deserve a brief comment. New contributions to the $\rho$ parameter stemming from Higgs states 
\cite{Rhoparam} have to comply with the current limits from precision measurements \cite{pdg}: 
$ |\delta\rho| \la 10^{-3}$ - there are limiting cases though, related to an underlying 
custodial symmetry,  where the extra contributions to $\delta\rho$ vanish. These limits are $m_h \simeq m_{H^\pm}$ and $\sin (\beta - \alpha) \simeq 0$, 
$m_H \simeq m_{H^\pm}$ and $\sin (\beta - \alpha) \simeq 1$, and finally $m_{H^\pm} \simeq m_A$. Values of $\tan \beta \lesssim 1$ together with a charged Higgs boson with a mass below 
a value of the order 100 $ GeV$ are disallowed by both the constraints coming from $R_b$ and from $B_q \bar{B_q}$ 
mixing~\cite{osland} for all Yukawa versions (also from the $b \to s \gamma$~\cite{bsgamma} measurement already discussed).   
In models Type II and Y,  the branching ratio of the process $B^+ \to \tau^+ \nu$ is enhanced relative to the SM. This is due to a $\tan^2 \beta$ factor, with origin in the Yukawa couplings, that is not present in the cases of models Type I and X~\cite{Logan:2009uf}. Therefore, no relevant bounds can be derived with this process for models Type I and X. Finally, the constraints that would arise from the precise measurements of the muon anomalous magnetic moment 
$(g-2)_\mu$  are irrelevant due to the large values of the masses involved in the process~\cite{g-2}.

\noindent
The theoretical bounds related to perturbative unitarity~\cite{unit1} 
and vacuum stability~\cite{vac1} (boundness from below) are also imposed and will prove to be 
very important in defining our benchmark scenarios. The most striking feature is that, as $\tan \beta$ grows,
the allowed values of $M$ shrink to a tiny region that depends mainly on $m_H$ and $\sin \alpha$.
This is extremely important because, as we will see later, when looking for the parameter space 
where a large enhancement of resonant cross sections is possible, one should be careful to definitely 
be in the parameter space region where perturbative unitarity and vacuum stability are respected. In figure 
\ref{fig:thconst} we present the allowed region for the 2HDM parameter space when theoretical constraints, 
vacuum stability and perturbative unitarity, are taken into account, in the plane   ($M^2$, $\sin \alpha$). 
We have chosen $\tan \beta = 8$ and $\tan \beta = 30$ and values for the masses allowed by all other 
experimental constraints. It is clear that, as $\tan \beta$ grows, the allowed parameter region shrinks.  
Inside this region of the parameter space, we can look for the largest possible enhancement of the cross sections.  
Finally, after choosing a CP-conserving minimum, the 2HDM vacuum is naturally protected against charge and CP 
breaking~\cite{charge}. There are no other bounds relevant for our analysis.

\section{$H^\pm$ production and decay channels at hadron colliders in 2HDM}
\label{Sec:Xsec+BRs}

\noindent
This section describes all relevant production and 
decay channels of charged Higgs bosons within 2HDMs at the 
LHC. Since a charged Higgs boson couples to fermions and bosons proportionally to their masses, 
it will predominantly be produced in connection with $\tau$'s, $b$'s and $t$'s as well as
$W^\pm$'s and $Z$'s or indeed in association with other scalars. Besides,
it will also decay into these states. As intimated, the discussion here will be focused 
on a light charged Higgs boson, that is, below the so-called threshold or transition region, 
when $m_{H^\pm}\approx m_t$. We will start by listing the most significant production modes which 
will then be discussed in detail in the following subsections. 
We can have single $H^\pm$ production  \cite{btH,bQH,cs1,cs2,cs3,cs4,WH,WHb,WHc,SH}:
\begin{subequations}\label{Eq:single-production}
\begin{align}
gg,q\bar q&\to b \bar t H^+ \text{ and } \bar b t H^-, \label{Eq:btH} \\
bQ        &\to bQ'H^+ \text{ and } bQ'H^-, \label{Eq:bQH}\\
cs        &\to H^{\pm}  (+jet), \label{Eq:qQH}\\
gg,b\bar b&\to W^-H^+ \text{ and } W^+H^-, \label{Eq:WH} \\
q Q &\to S H^+ \text{ and } S H^-, \label{Eq:SH} 
\end{align}
\end{subequations}
where $S$ is a neutral scalar ($S=h,H,A$), or pair production \cite{HpHm1,HpHm2,DY,VBF} (see also \cite{HH}):
\begin{subequations}\label{Eq:pair-production}
\begin{align}
gg,b\bar b&\to H^+H^-, \label{Eq:HH} \\
q\bar q   &\to H^+H^-, \label{Eq:DY} \\
qQ        &\to q'Q' H^+H^-, \label{Eq:VBF}
\end{align}
\end{subequations}
where $q,q^\prime,Q,Q^\prime$ represent (anti)quarks (other than $b$'s and $t$'s). Notice that process 
(\ref{Eq:btH}) contains as subprocesses both top-antitop production followed by the decay
$\bar t\to \bar b H^-$ (and c.c.) and $bg \to t H^-$ (and c.c.): see, e.g., \cite{threshold}. 
The process $q Q \to S H^+ \text{ and } S H^-$ requires a more elaborate analysis because the decays 
of the scalar that is produced together with the charged Higgs boson is Yukawa Type model 
dependent~\cite{Arhrib:2009hc, Kanemura:2009mk}. A complete study of this process is in progress and will be 
presented elsewhere~\cite{us1}.

\noindent
Although the charged Higgs boson has a number of possible decay modes, in models Type I and X and 
variations thereof, a light charged Higgs boson decays mainly via $H^+ \to \tau^+ \nu$ and $H^+ \to c \bar{s}$. 
The BR of each channel depends exclusively on the charged Higgs boson mass and on 
$\tan \beta$ as long as the decays to neutral scalars and a $W^\pm$ boson are kinematically  forbidden. 
It is well known that a very light (neutral) CP-even Higgs state is still allowed in the context of multi-Higgs doublets. 
In 2HDMs, it suffices to take a very small value of $\sin (\beta - \alpha)$~\cite{Kalinowski:1995dm} to avoid 
the LEP SM Higgs bound~\cite{Schael:2006cr}. In this work we will consider the mass of the lightest CP-even 
Higgs boson to be above 100 $GeV$, disallowing therefore the decay $H^+ \to W^+ h$. The mass of the remaining 
neutral scalars are also considered to be above 100 $GeV$. For each production process discussed, we will briefly 
comment on the very light CP-even Higgs scenario, if pertinent.

\begin{figure}[h!]
\centering
\includegraphics[height=3.05in]{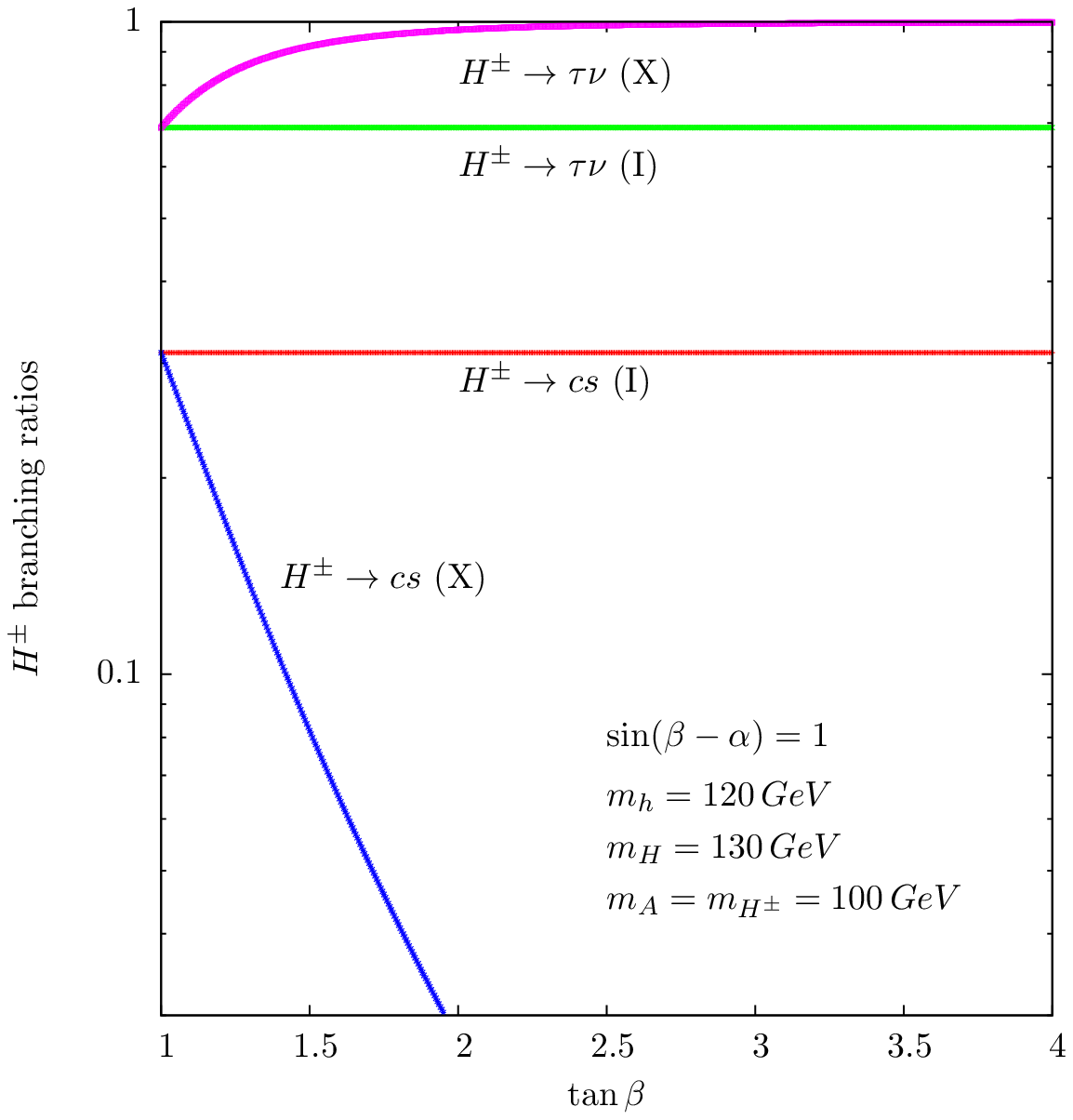}
\hskip -.2cm
\includegraphics[height=3.05in]{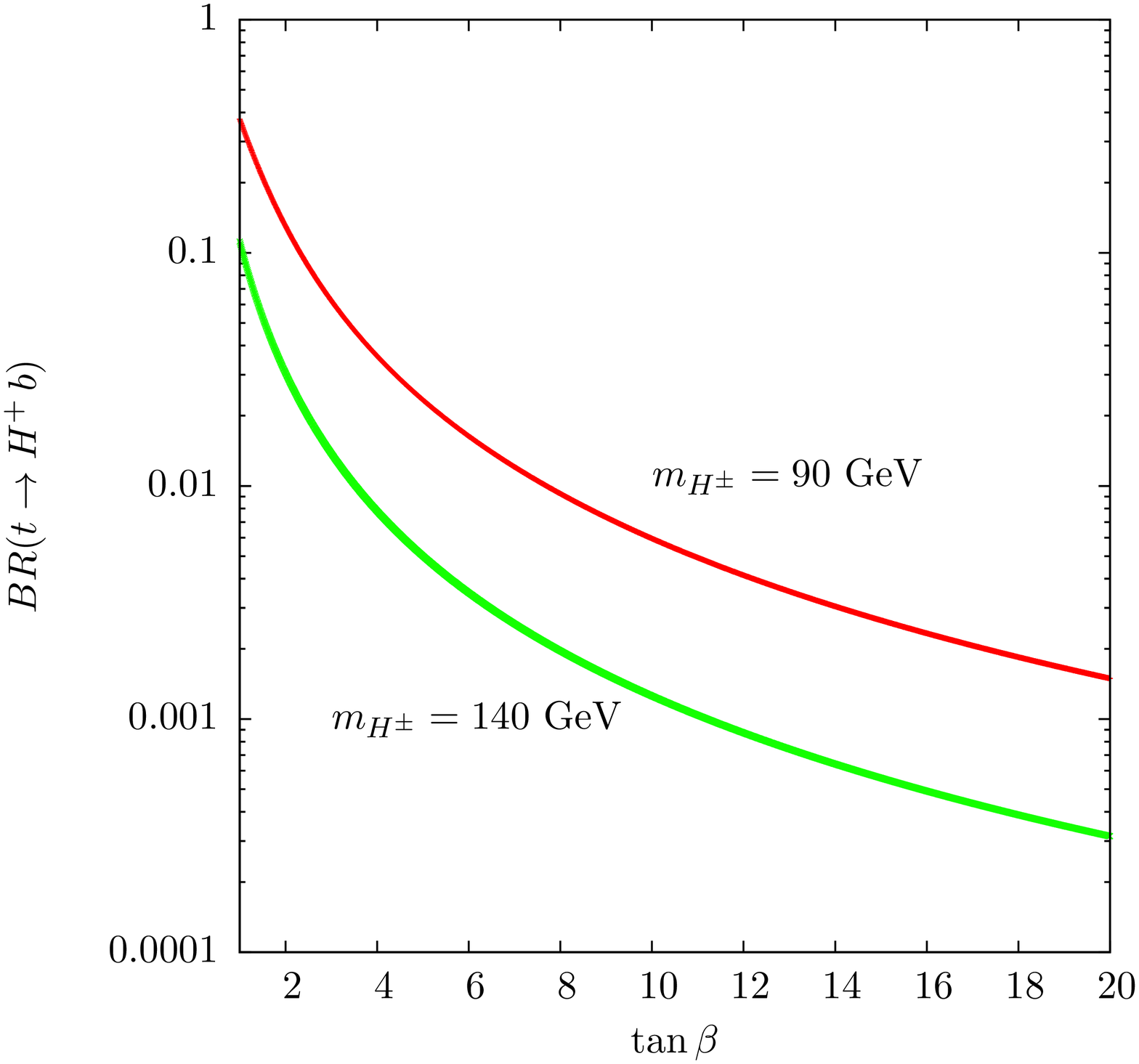}
\caption {In the left panel we present the charged Higgs BRs for $m_{H^\pm}=100$ $ GeV$ as a function of 
$\tan \beta$ in models Type I and X. In the right panel 
we show BR$(t \to H^+ b)$ as a function of $\tan \beta$ for two values of the charged Higgs boson mass.}
\label{fig:BR}
\end{figure}
\vskip -0.1cm

\noindent
In the left panel of figure~\ref{fig:BR} we present the BRs for $m_{H^\pm}=100$ $ GeV$ as a function of $\tan \beta$ in 
models Type I and X (the values of the remaining parameters are shown in the figure). It is clear that, in 
this type of models, $H^+ \to \tau^+ \nu$ is by far the dominant decay mode. Because the charged Higgs boson width 
depends only on $\tan \beta$ and on the charged Higgs boson mass, the plot is representative of all 
$m_{H^\pm}$'s values below the $t\bar b$ threshold provided no light neutral scalars are present. 
At the LHC, the most promising search mode for a light charged Higgs boson is $pp \to \bar t t$ with subsequent 
decays to $\bar b b W^\pm H^\mp$. Therefore, to understand how the (anti)top decays change relative to the SM, where 
the BR$(t\to W^+ b)$
is close to 100\%,  we show in the right panel of the figure the BR$(t \to H^+ b)$ as a function of 
$\tan \beta$ for two values of the charged Higgs boson mass. Contrary to the case of 
the MSSM and MSSM-like versions of a Type II 2HDM, this BR falls very rapidly with $\tan \beta$ here, 
which makes this mode useless for values of $\tan \beta \gtrsim 10$ and even  more so as the charged Higgs 
boson mass approaches the top quark one. Therefore, we conclude that in these models light charged Higgs 
bosons decay predominantly to $\tau \nu$ and that searches based on (anti)top-quark decays to charged Higgs 
bosons are only effective for low to moderate values of $\tan \beta$.  We will now discuss each production 
mode in turn.  


\subsection{Double top production}
\label{sec:ttbar}

\begin{figure}[h]
\centering
\includegraphics[height=3.1in]{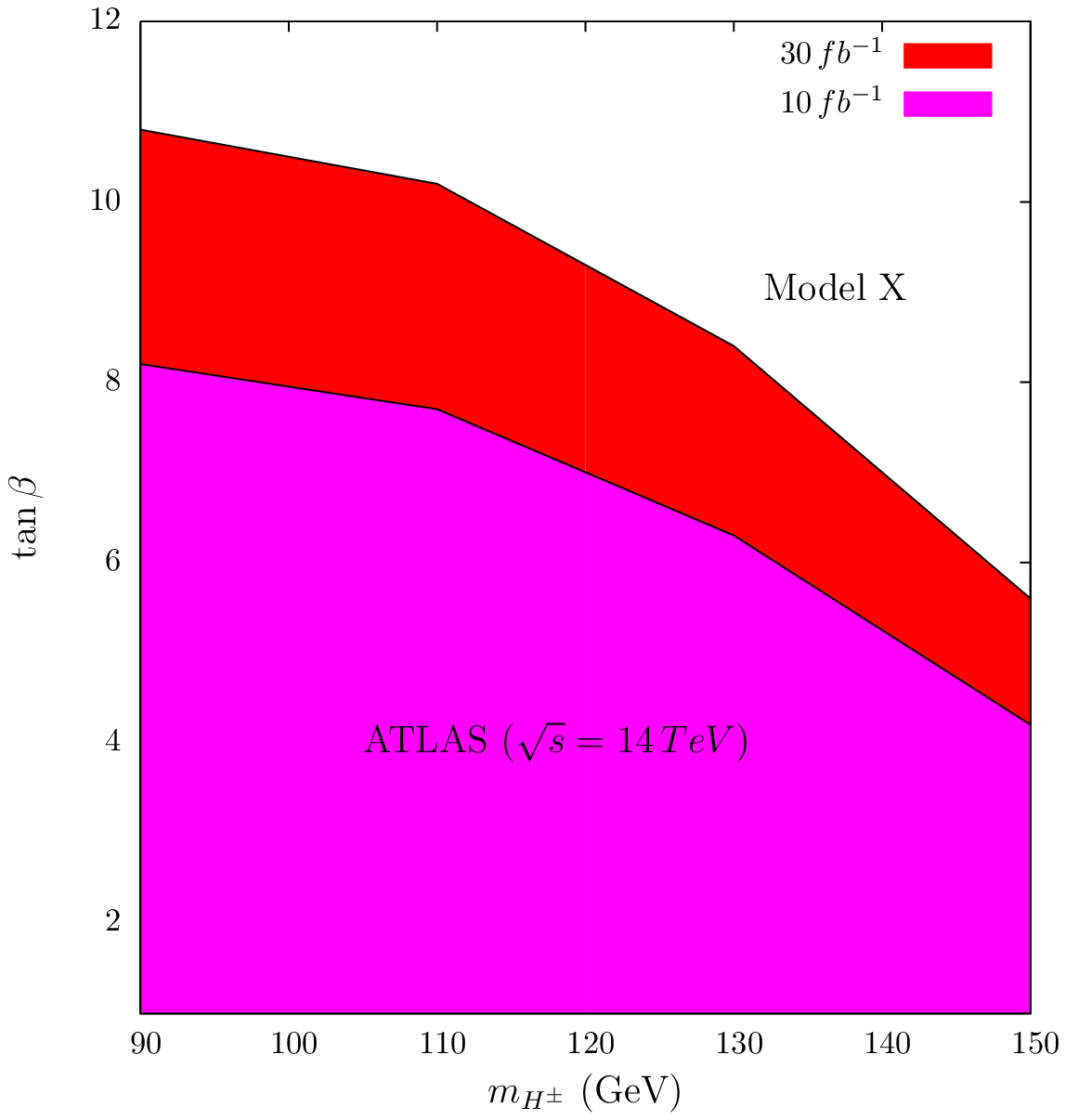}
\includegraphics[height=3.1in]{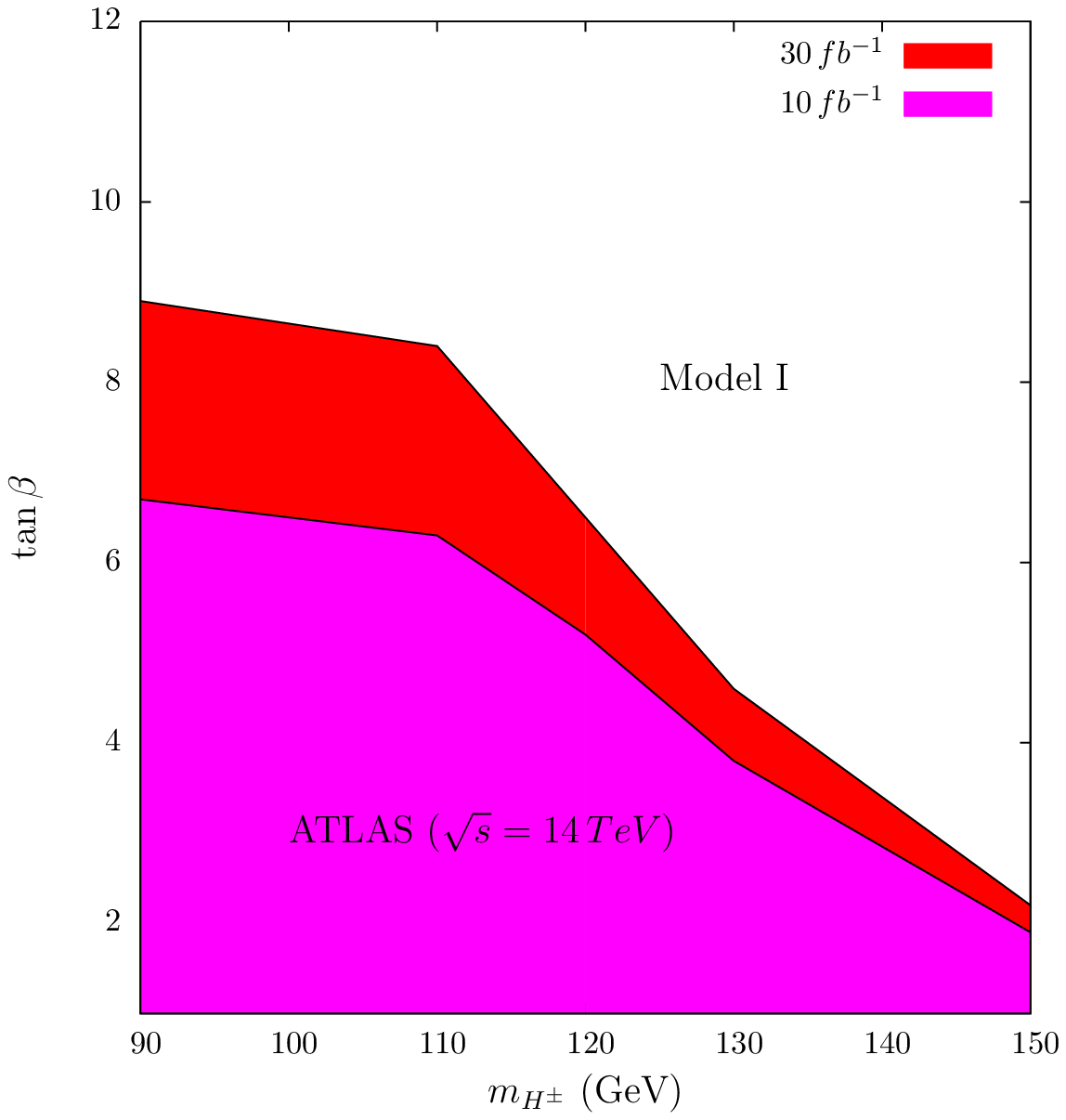}
\caption {Region of the parameter space excluded by the ATLAS collaboration for 
$\sqrt{s} = 14 \, TeV$ in models Type X (left) and Type I (right) using the channel 
$pp \to t \bar{t} \to H^{\pm} b W^{\pm} \bar{b} \to \tau \nu b \bar{b} q \bar{q}$. The plot
was done using the data in~\cite{ATLAS}.}
\label{fig:ATLAS}
\end{figure}
\vskip -0.3cm

\noindent
The most promising searches for a light charged Higgs at the LHC were performed in the 
$pp \to t \bar{t} \to H^{\pm} b W^{\pm} \bar{b} \to \tau \nu b \bar{b} q \bar{q}$ channel. 
This is the first process on our list (\ref{Eq:btH}) where both top and anti-top are produced on-shell. 
This production mode depends only on $\tan \beta$ and on the charged Higgs mass via both the top and the 
charged Higgs boson BRs. Therefore, the region of parameter space probed can be shown in the 
$(\tan \beta, m_{H^\pm})$ plane and is independent of all other 2HDM parameters. Both the ATLAS~\cite{ATLAS} 
and the CMS~\cite{CMS} collaborations have performed full detector analyses in this channel and have used them to explore the 
parameter space of the MSSM. We can now use their raw results to constrain the parameter space of 2HDMs 
Type I and X. In figure~\ref{fig:ATLAS} we present the region of the parameter space that can be excluded at 
95\% CL in model Type X (left panel) and Type I (right panel) at the LHC after collecting 10 and 30 
$fb^{-1}$ of integrated luminosity using the results from the ATLAS collaboration~\cite{ATLAS}.
Very briefly, the procedure to extract  the data from~\cite{ATLAS} was the following. 
The ATLAS collaboration has generated the signal $pp \to t \bar{t} \to H^{\pm} b W^{\pm} \bar{b} \to \tau \nu b \bar{b} q \bar{q}$ 
using MSSM tools that are only relevant in the calculation of
$BR (t \to H^{\pm} b )$ and $BR (H^{\pm} \to \tau \nu )$ - 
all the remaining cross sections and branching ratios are SM like (including all the backgrounds). 
We then took the final values in~\cite{ATLAS} for a given point in parameter space, 
divided them by the MSSM values for $BR (t \to H^{\pm} b )$ and $BR (H^{\pm} \to \tau \nu )$ 
and multiply them for the corresponding values for the 2HDM model under study.

\noindent
Contrary to the MSSM scenario described by the ATLAS collaboration in~\cite{ATLAS}, the 2HDM parameter space will not be completely covered with a 
luminosity of $30 \, fb^{-1}$. As previously discussed, 2HDMs Type I and X have Yukawa couplings 
proportional to $1/\tan \beta$. Hence, contrary to the MSSM where a term $m_b \, \tan \beta$ 
(where $m_b$ is the bottom-quark mass)  gives large contributions for large $\tan \beta$, in Types I and X 
this term in now $m_b / \tan \beta$, thus it decreases with $\tan \beta$. Moreover,  an increase in luminosity 
will make the excluded region to grow to larger values of $\tan \beta$, but because BR$(t \to \bar{b} H^+)$ 
always decreases as $\tan \beta$ grows, very large values of $\tan \beta$ will never be probed with this process.


\subsection{Single top production}

\noindent
In this section we estimate the contribution of the process $pp\rightarrow H^\pm b j$ with 
$H^\pm \to \tau \nu_\tau $ to light charged Higgs boson searches at the LHC in 2HDMs. This mode consists 
mainly of  $t$-channel plus $s$-channel single top production followed by the decay $t \to H^+ \bar b$. 
Our goal is to understand whether this process, that again only depends on the charged 
Higgs boson mass and $\tan \beta$,  could contribute to improve the region already scrutinised by 
$pp \to t \bar{t} \to H^{\pm} b W^{\pm} \bar{b} \to \tau \nu b \bar{b} q \bar{q}$ presented in the previous 
section. A first study for this process was presented in~\cite{bQH}. To estimate single top 
contribution, we take as our signal the process
\begin{equation}
pp\rightarrow H^\pm b j\rightarrow b j\tau \nu_\tau.
\end{equation}
Regarding the background, we consider the irreducible background process
\begin{equation}
pp\rightarrow b j l(\tau,\mu,e) \nu_{\tau,\mu,e}
\end{equation}
and the reducible one
\begin{equation}
pp\rightarrow t\bar t\rightarrow W^+ b W^-\bar b,
\end{equation}
where both $W^\pm$'s can decay semi-leptonically, fully leptonically or fully hadronically. The details of the 
parton level analysis are presented in Appendix A. For the signal, we have varied the charged Higgs boson mass from  
$m_{H^\pm}=100$  $ GeV$ to $m_{H^\pm}=140$  $ GeV$ in 10 $ GeV$ steps and the results are for $\tan\beta=1.5$. The signal 
was generated with CalcHEP~\cite{Pukhov:2004ca} and so was the $t \bar t$ background. The single top background, 
$pp\rightarrow b j l(\tau,\mu,e) \nu_{\tau,\mu,e}$, was instead generated with 
MadGraph/MadEvent~\cite{Alwall:2007st}. The partonic CM energy was chosen as both the renormalisation and factorisation scale. 

\begin{table}[h!]
\begin{tabular}{cccccc}
\hline
  & {${m_{H^\pm}=100 \, {GeV}}$}  & {${m_{H^\pm}=110 \, {GeV}}$}  & {${m_{H^\pm}=120 \, {GeV}}$}  & { 
${m_{H^\pm}=130 \, {GeV}}$} & { ${m_{H^\pm}=140 \, {GeV}}$} \\ \hline \hline
 {Process}  &   \\
\hline
Signal & 379.4 $fb$ & 274.4 $fb$  & 202.7 $fb$  & 118.9 $fb$  & 65.5 $fb$ \\
\hline
Bg (single top) & 1705.4 $fb$ &   &   &   & \\
Bg ($t\bar t$ semi-leptonic) & 683.1 $fb$ &   &   &   & \\
Bg ($t\bar t$ leptonic) & 393.6 $fb$ &   &   &   & \\
\hline
$\sigma_S/\sigma_B$ & 0.14 & 0.098  & 0.073  & 0.042  & 0.023 \\
\hline
$\sigma_S/\sqrt{\sigma_B}$  ($fb^{1/2}$) & 7.19 & 5.20  & 3.84  & 2.25  & 1.24 \\
\hline
\end{tabular}
\caption{Cross section for the signal $pp\rightarrow H^\pm b j\rightarrow b j\tau \nu_\tau$ and for the irreducible 
($pp\rightarrow b j l(\tau,\mu,e) 
\nu_{\tau,\mu,e}$) and
main reducible ($pp\rightarrow t\bar t\rightarrow W^+ b W^-\bar b$) backgrounds after all cuts have been applied. 
The analysis is performed for model Type X with $\tan \beta = 1.5$ and the single top process has a negligible 
dependence on the remaining parameters of the 2HDM. (Background rates are independent of $m_{H^\pm}$ as the
latter parameter is not used in the selection, see Appendix A.)}
\label{tab:singtop}
\end{table}

\noindent
By performing the analysis presented in Appendix A we have found the results presented in table \ref{tab:singtop}. Our results show that, 
for a mass of 100 $GeV$, an exclusion of  $\tan \beta \lesssim 5$ could be expected at 95\% CL for a collected luminosity of $10 \, fb^{-1}$ 
and $\tan \beta \lesssim 7$ for $30 \, fb^{-1}$, for model Type X. The results then degrade rapidly with 
growing charged Higgs boson mass as compared 
to the previous process of $t \bar t$ production, e.g, for a 140 $GeV$ $H^\pm$ mass, even a value of $\tan \beta = 3$ is not reachable 
with 30 $fb^{-1}$ of integrated luminosity at 95\% CL. As discussed in Appendix A, these results will most probably prove to be optimistic as this is a parton 
level analysis. However, we believe to have made a case for the need of further studies of this process and we do think that a full detector level 
analysis would be worth performing. Finally, in accordance with the previous section, the results are slightly worse for model Type I due 
to the reduction of the decay rate to $\tau \nu$ when compared to model Type X.


\subsection{Direct charged Higgs boson production}

\begin{figure}[h!]
\centering
\includegraphics[height=1.2in]{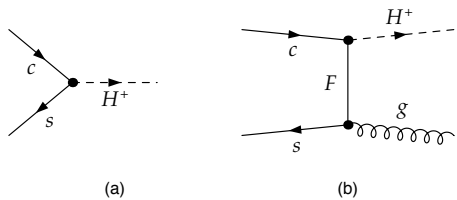}
\caption {Representative Feynman diagrams for direct charged Higgs boson production and charged Higgs boson plus jet 
production. (Herein, $F$ represents a fermion, as appropriate.)}
\label{fig:proc2c}
\end{figure}
\vskip -0.0cm

\noindent
The last process on our list that depends only on the charged Higgs boson mass and on $\tan \beta$ is direct 
charged Higgs boson production 
$cs  \to H^{\pm} ~ (+~jet)$ and is shown in figure~\ref{fig:proc2c}. As discussed before, contrary to the MSSM and 2HDMs Type II and Y, in models Type I and X,  the couplings of the charged Higgs to both up- and 
down-type quarks is proportional to $1/\tan \beta$. This is the reason why we write as initial state only $cs$ and not $cb + cs$ - 
the $cb$ initiated mode can be the dominant one for direct production in the above models, as discussed in~\cite{cs1,cs2}. In models Type I and 
X the ratio between the $cb$ and the $cs$ initiated contributions is of the order $V^2_{cb} \, (m^2_c + m^2_b)/m^2_c$ which makes the $cb$ contribution 
negligible for all charged Higgs masses and for all $\tan \beta$ values. 

\noindent
A detailed experimental study for direct charged Higgs boson production with subsequent decay to $\tau \nu$ was performed by CMS~\cite{cs3}. In that study, 
it was shown that values of $\tan \beta$ above 15 could be excluded at 99\% CL for $m_{H^\pm}=200$ $ GeV$ for the MSSM. According to~\cite{cs2}, 
this corresponds to a cross section of the order of a few hundreds of $fb$.  However, for models Type I and X, the cross sections are below 1 
$fb$ even for $\tan \beta =1$, where they reach their maximum value. If we take the lower limit for the mass,  $m_{H^\pm}=100$ $GeV$, then the 
typical cross section is of the order of 700 $fb$ for $\tan \beta = 1$. This seems like a large value but there are two drawbacks. First, the value 
$m_{H^\pm}=100~GeV$ is close to the $W^\pm$ mass and therefore the irreducible background will make the detection much harder. Second, as soon as we move 
to, say, $\tan \beta = 4$ the cross section is reduced to values of the order of 50 $fb$. Hence, we conclude that this process will not give any 
significant contribution to the limits in the $(m_{H^\pm}, \tan \beta)$ plane established in the previous sections for 2HDM Type I and X.


\subsection{Charged Higgs boson pair production}

\begin{figure}[h!]
\centering
\includegraphics[height=2.4in]{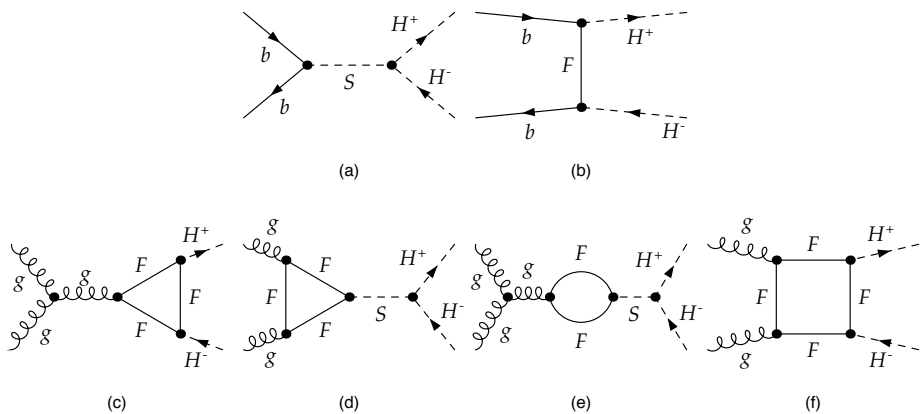}
\caption {Representative Feynman diagrams for charged Higgs boson pair production via $b\bar b$, (a)-(b), and 
$gg$, (c)-(f), fusion. (Herein, $F$ represents a fermion, as appropriate, whereas $S$ a neutral Higgs boson,
as appropriate.)}
\label{fig:proc3a}
\end{figure}

\begin{figure}[h]
\centering
\includegraphics[height=0.9in]{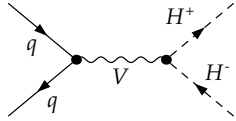}
\caption {Feynman diagrams for DY charged Higgs boson pair production. 
(Herein, $V$ represents a neutral gauge boson, as appropriate.)}
\label{fig:proc3b}
\end{figure}

\noindent
Charged Higgs pair production in 2HDMs at the LHC proceeds via three different channels, $gg \to  H^+ H^-$,  $b \bar{b}  \to  H^+ H^-$ and the 
Drell-Yan (DY) process $q \bar{q}  \to  H^+ H^-$, where $q$ stands for a light quark. The first two processes are 
presented in figure~\ref{fig:proc3a} whilst 
the last one is shown in figure ~\ref{fig:proc3b}.  The most recent study that compares all three processes in the MSSM was performed 
in~\cite{HpHm2}. Therein, it was shown that, for the MSSM, the main contribution comes from DY except for very large values of $\tan \beta$ 
(above 50) where gluon fusion gives the main contribution. However, when we consider 2HDMs of Type I and X, there is never a $\tan \beta$ 
enhancement related to the Yukawa couplings of charged Higgs bosons to quarks as these are always proportional to $1/\tan \beta$. 
Conversely, because in 2HDMs we are not constrained by relations between the scalar masses, large cross sections can be obtained if we consider resonant 
production of neutral Higgs bosons decaying into pairs of charged ones. In this scenario, cross sections of $pb$ order  can be reached if one 
considers a further enhancement due to Higgs self-couplings. In this scenario, the $b \bar b \to H^+ H^-$ mode (diagrams (a) and (b)
in figure \ref{fig:proc3a}) is always 
much smaller than the gluon fusion mode
(diagrams (c)--(f) in figure \ref{fig:proc3a}), especially for large $\tan \beta$, because the Yukawa enhancement only happens in 2HDM Type II, Type Y and in 
the MSSM.  As it has been noted that the large enhancement to the production cross section can only happen in resonant production, it 
is always diagram (d) (and partly (a)) in figure~\ref{fig:proc3a} that dominates in these circumstances, through the exchange of a heavy CP-even Higgs. 

\vskip -0.1cm

\vskip -0.1cm

\noindent
We showed in previous sections that, for a charged Higgs boson mass of 100 $GeV$, the processes that depend only on $\tan \beta$ and on the 
charged Higgs boson mass itself will at least exclude a region where $\tan \beta \lesssim 8$ (e.g., in model Type X for 10 $fb^{-1}$ of integrated 
luminosity and $\sqrt{s} = 14$ $TeV$). The situation could even be improved when a thorough study for single top production is combined with the one 
already performed for the $t \bar{t}$ case by the ATLAS and the CMS collaborations. As the charged Higgs mass grows, the excluded range of $\tan \beta$ shrinks: e.g.,
for a 150 $ GeV$ charged Higgs boson mass we get $\tan \beta \gtrsim 2$ for model Type I and $\tan \beta \gtrsim 4$ for model Type X, 
when 10 $fb^{-1}$ are collected. Therefore, we want to see whether large values of  $\tan \beta$ can give significant cross sections in charged Higgs 
boson pair production. This would allow us to look into the high $\tan \beta$ region which we already know will never be accessible with the previous 
reactions (3a, 3b and 3c). However, we first have to understand what can be considered a significant cross section. For this purpose we 
make use of the studies from~\cite{Datta:1999nc}  made in the context of left-right symmetric models
and~\cite{Davidson:2010sf} performed in the context of a 2HDM-like model, where three 
gauge-singlet right-handed Weyl spinors were added to become the right-handed components of the three Dirac neutrinos. Using these studies~\cite{Datta:1999nc,Davidson:2010sf} as 
a guide, we conclude that a significant cross section is one of the order of 400 $fb$. This means that, using their analysis for our signal one would be able to start probing a significant 
parameter region with 30 $fb^{-1}$ for 2HDM Type X if the production cross section was of the order of 400 $fb$. The details of the analysis are presented in Appendix B.

\begin{figure}[h]
\centering
\includegraphics[height=3.15in]{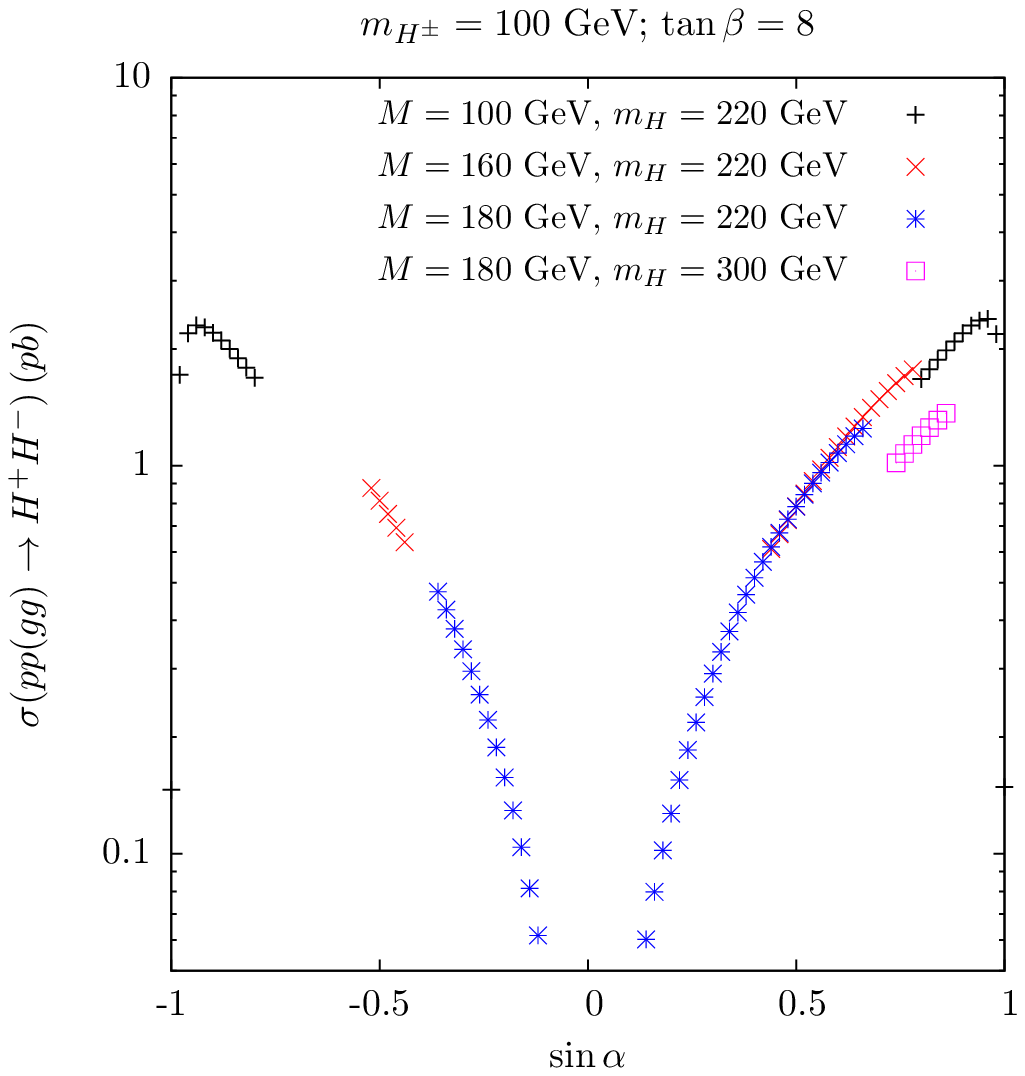}
\includegraphics[height=3.15in]{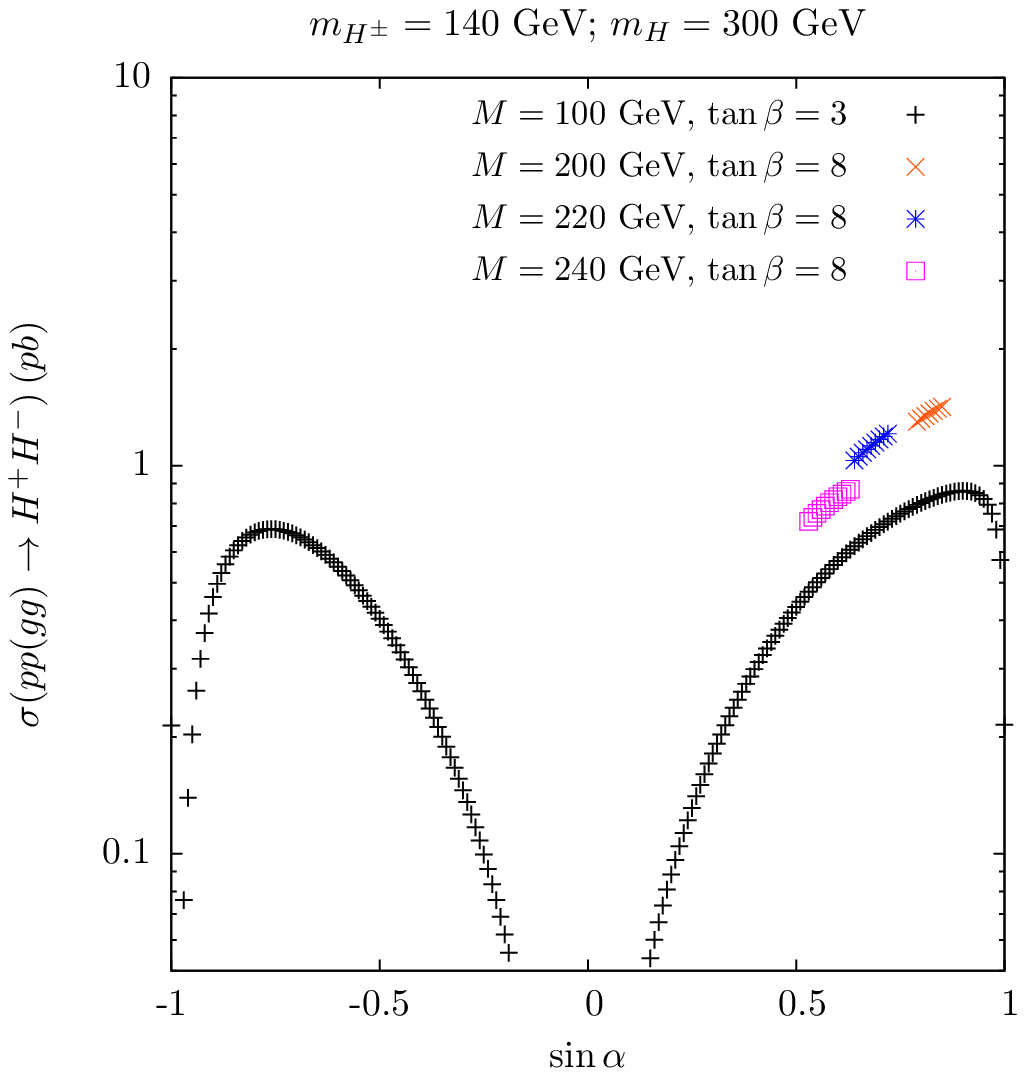}
\caption {Production cross section for $pp (gg) \to H^+ H^-$ as a function of $\sin \alpha$ for $m_{H^\pm} =100$ $ GeV$ (left) and $m_{H^\pm} =140$ $ GeV$ 
(right) for a set of chosen values of $\tan \beta$, $m_{H}$ and $M$.}
\label{fig:ggHH1}
\end{figure}
\vskip -0.1cm

\noindent
Before showing our results for charged Higgs pair production, let us first analyse the high $\tan \beta$ behaviour of the cross section. If, in fact, no $\tan \beta$ enhancement can be expected for this process, it 
will be useless because the previous processes 3a, 3b and 3c already constrain a significant region of the $(\tan \beta, m_{H^\pm})$ plane. It is easy to show that for high $\tan \beta$ we have
\begin{equation}
g_{H H^+ H^-} \, g_{H \bar t t}   \,  \propto \, \sin \alpha \, \cos \alpha \, \tan \beta \, (m_H^2 - M^2)  \qquad (\tan \beta \gg 1)
\end{equation}
where $g_{H H^+ H^-}$ is the coupling of the heavy CP-even Higgs to a charged Higgs pair while $g_{H \bar t t} $ is the $H$ Yukawa coupling to top quarks. Therefore, it becomes clear that the cross section increases with $\tan^2 \beta$ and can be used to explore the high  $\tan \beta$ region. Asymptotically, there is also a term independent of $\tan \beta$ of the form $\sin^2 \alpha \,  (2 \,m_{H^\pm}^2 - M^2)$ that can be important for moderate values of $\tan \beta$. It is interesting to note that the lightest CP-even Higgs state
 has a similar behaviour in the same regime
\begin{equation}
g_{h H^+ H^-} \, g_{h \bar t t}   \,  \propto \, \sin \alpha \, \cos \alpha \, \tan \beta \, (m_h^2 - M^2)  \qquad (\tan \beta \gg 1)
\end{equation}
where the couplings refer now to the lightest CP-even Higgs state. Hence, if both $H$ and $h$ have masses above two times the charged Higgs boson mass,
 their contributions will be indistinguishable at high $\tan \beta$. In the left panel of figure~\ref{fig:ggHH1} we present the cross section for $gg \to H^+ H^-$ with $m_{H^\pm} =100$ $GeV$, $\tan \beta = 8$ and for 
several values of $(M, m_H)$ as a function of $\sin \alpha$. Some lines on the plots end or/and start abruptly due to the theoretical constraints:
this means that these are constraints that will not change with time and can be regarded as a part of the parameter space that is definitely excluded. 
It is clear that large cross sections can be obtained for various regions of the parameter space as long as the production is resonant. In the right 
panel we show the cross section for $m_{H^\pm} =140$ $ GeV$ and for a fixed $m_H$ of 300 $ GeV$ for several values of $(M, \tan \beta)$ as a function of $
\sin \alpha$. Again, we see that values of the cross section close to 1 $pb$ can easily be obtained. The light CP-even Higgs mass was taken to be 
120 $ GeV$ while the CP-odd boson mass $m_A $ was chosen so as to avoid the $\delta \rho$ constraint.  The calculations were performed with the packages 
FeynArts~\cite{feynarts} and FormCalc~\cite{formcalc}. The scalar integrals were evaluated with  LoopTools~\cite{looptools} and the CTEQ6L parton 
distribution functions~\cite{cteq} were used. Again, the partonic CM energy was chosen as both the factorisation and renormalisation scale. The value of 
the DY cross section was not taken into account. This means that taking a charged Higgs mass of 100 $ GeV$ a value of the order of  200 $fb$ should be 
added to the values presented in the plot. We have repeated the calculation for the same processes but initiated by $b \bar{b}$ and found that the cross 
sections are usually at least one order of magnitude below the $gg$ initiated process.

\begin{figure}[h]
\centering
\includegraphics[height=3.6in]{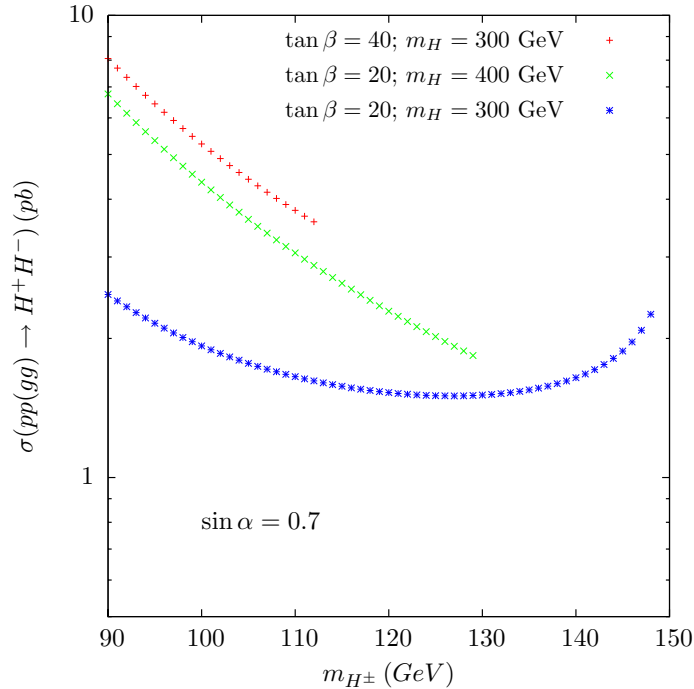}
\caption {Production cross section for $pp (gg) \to H^+ H^-$ as a function of the charged Higgs boson mass. The values of $\sin \alpha$ were chosen 
so as to maximise the cross section while complying with all the constraints.}
\label{fig:ggHH2}
\end{figure}
\vskip -0.1cm

\noindent
In figure \ref{fig:ggHH2} we present the total cross section for $pp (gg) \to H^+ H^-$  as a function of the charged Higgs boson mass. The values of 
$\tan \beta$ and $\sin \alpha$ were chosen so as to maximise the cross section while in agreement with all constraints. These are the maximum values the 
cross section can attain, a few $pb$, which is not surprising because the production cross section via $gg$ fusion for a SM Higgs boson is of the order of a few  
tenths of $pb$. Even considering the decay into a pair of charged Higgs bosons to be close to 100\%, the final cross section values cannot exceed a 
few $pb$. We note that this process is highly dependent on the available channels for the $H$ state to decay into, though. If, for example, we would close 
the channel $H \to hh$ by increasing the lightest CP-even Higgs boson mass, the cross section would become larger in some regions of the parameter space. We 
should however note the importance of this process in probing the large $\tan \beta$ region. As in these models the charged Higgs Yukawa couplings decrease like 
$1/ \tan \beta$, only the self-coupling $H H^+ H^-$ allows for large values of the cross section for high $\tan \beta$. 
Finally, a word about the very 
light CP-even Higgs scenario is in order. In the large $\tan \beta$ regime, $\sin (\beta - \alpha) \approx \cos \alpha$. 
The largest values of the cross section are obtained for $\sin \alpha \approx 0.7$, which means $\sin (\beta - \alpha) 
\approx  \cos \alpha \approx 0.7$. Consequently, the LEP bound forces the lightest CP-even Higgs mass to be above 100 $GeV$. However, we can still 
get sizeable cross sections for $\sin \alpha$ very close to 1. In that case, not only a very light CP-even Higgs boson is allowed but, moreover,
the decay $H^+ \to W^+ h$ becomes maximal, as it is proportional to  $\cos (\beta - \alpha)$. 

\noindent
We end this section with a brief comment on the cross section values for $M^2 < 0$. First, note that the values of $M^2$ affect only the processes we define in this work as charged Higgs pair production and Vector Boson Fusion (VBF), via the vertices $H (h) H^+ H^-$. As discussed in~\cite{Cornet:2008nq, Arhrib:2009gg} , negative values of $M^2$ can give rise to very large values of the cross section of neutral Higgs pair production as a result of an enhancement of the $H (h) H^+ H^-$ self-couplings. The same is true for charged Higgs pair production and VBF. We have previously shown the asymptotic behaviour of the $H H^+ H^-$ coupling, $g_{H H^+ H^-} \, g_{H \bar t t}   \,  \propto \, \sin \alpha \, \cos \alpha \, \tan \beta \, (m_H^2 - M^2)  \qquad (\tan \beta \gg 1)$. Therefore, it is clear that the largest values of $g_{H H^+ H^-} \, g_{H \bar t t}$ are obtained for $M^2 < 0$. However, we have shown in~\cite{Belyaev:2009zd} that, when $M^2 < 0$, $\tan \beta$ is constrained to be small by perturbative unitarity and more so if $|M^2|$ becomes very large. We have shown in~\cite{Belyaev:2009zd} that it is very hard to go beyond a value of $\tan \beta \approx 8$ even for $m_H=130 \, GeV$ - the larger $m_H$ is the more constrained $\tan \beta$ is, independently of the value of  $M^2 < 0$. Hence, as the range of small to moderate $\tan \beta$ will be covered by the processes that depend only on $\tan \beta$ and on the charged Higgs mass, there is no need for a detailed study of the $M^2 < 0$ scenario.


\subsection{Vector Boson Fusion}

\begin{figure}[h!]
\centering
\includegraphics[height=1.2in]{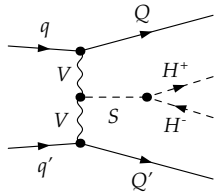}
\caption {Feynman diagram for resonant charged Higgs boson production via VBF. (Herein, $S$ represents a neutral Higgs boson,
as appropriate.) The complete set of 
Feynman diagrams for VBF can be found in \cite{VBF}. However large cross sections are obtained just for resonant production.}
\label{fig:proc3c}
\end{figure}
\vskip -0.0cm

\noindent
In the SM, VBF is the closest competitor to Higgs boson production via gluon fusion,
in terms of inclusive rates. Although it presents a smaller value for the the total cross section, it has obvious advantages in the analysis 
due the background reduction accomplished by using the two very forward/backward jets that  accompany  the Higgs boson
in the final state. In the case of 2HDMs and heavier charged Higgs bosons, VBF can be induced at one-loop level for the case of singly produced
$H^\pm$ states (see \cite{Kanemura:2000si,Kanemura:2000cw} for the case of $e^+e^-$ colliders), however, we have 
verified here that the corresponding hadro-production rates are negligible.
\begin{figure}[h!]
\vskip 1.cm
\centering
\includegraphics[height=3.6in]{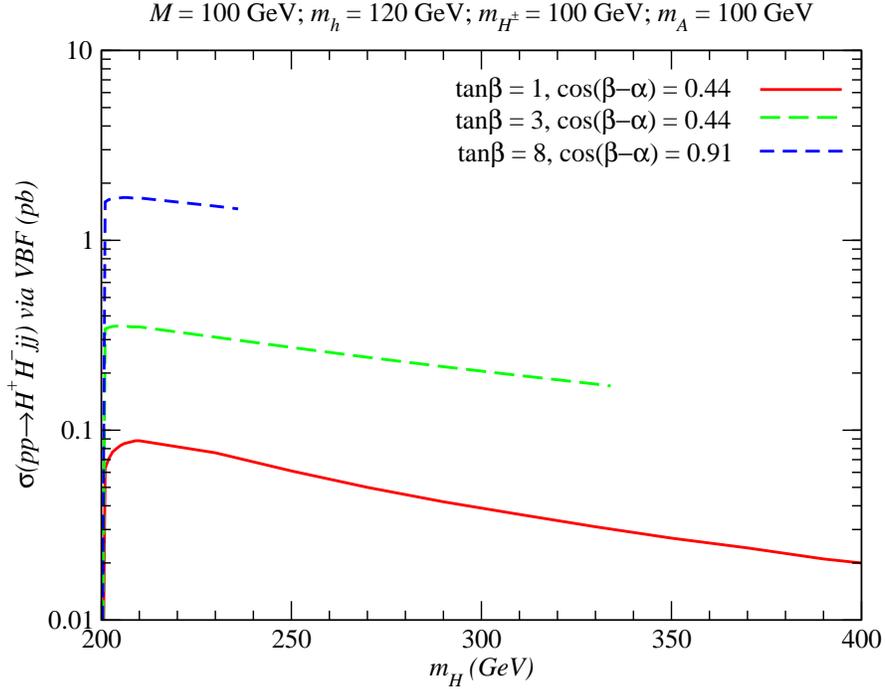}
\caption {Production cross section for $pp \to H^+ H^- jj$ via VBF as a function of the heaviest CP-even Higgs boson mass. The values of $\sin \alpha$ were chosen 
so as to maximise the cross section while complying with all available constraints.}
\label{fig:wwh}
\end{figure}
\vskip -0.cm
Production of $H^\pm$ pairs is instead possible at tree level from VBF and,
like with the previous production process, only the resonant diagram in figure \ref{fig:proc3c} together with an enhanced $H H^+ H^-$ coupling gives rise 
to sizeable values of the total cross section. A detailed parton level study for $pp \to  H^+ H^- jj$ was performed for VBF in the context of the 
MSSM in~\cite{VBF}. This study considered only  charged Higgs masses above the $tb$ threshold and therefore the largest number of signal events was 
obtained for the final state  $H^+ H^- jj \to b \bar b W^+ \tau^ - \nu jj \to b \bar b jjjj \tau \nu \to 6j + \tau + \slashed{p}_T$. Therein it was shown 
that in the framework of the MSSM and for the LHC ($\sqrt{s}=14$ $TeV$) this final state will be approximately 1000 times below the QCD background. 
Clearly, below the $tb$ threshold, the largest number of signal events is obtained with  final states $H^+ H^- jj \to  2\tau + 2j + \slashed{p}_T$ 
instead. This is also a cleaner detection mode for which there is no available study in the literature. However, using the SM processes as a guide we 
would conclude that cross sections close to the $pb$ level will most certainly be accessible at the LHC.

\begin{figure}[h!]
\centering
\includegraphics[height=3.15in]{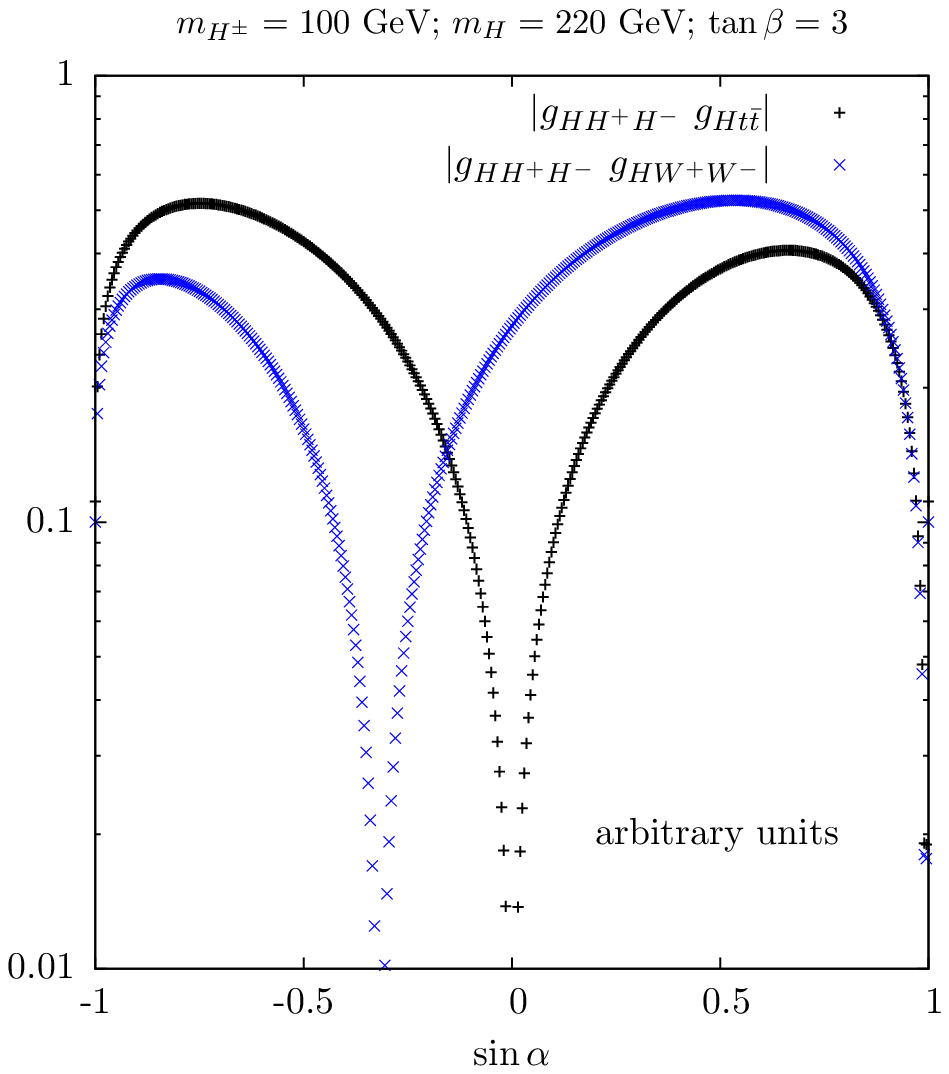}
\includegraphics[height=3.15in]{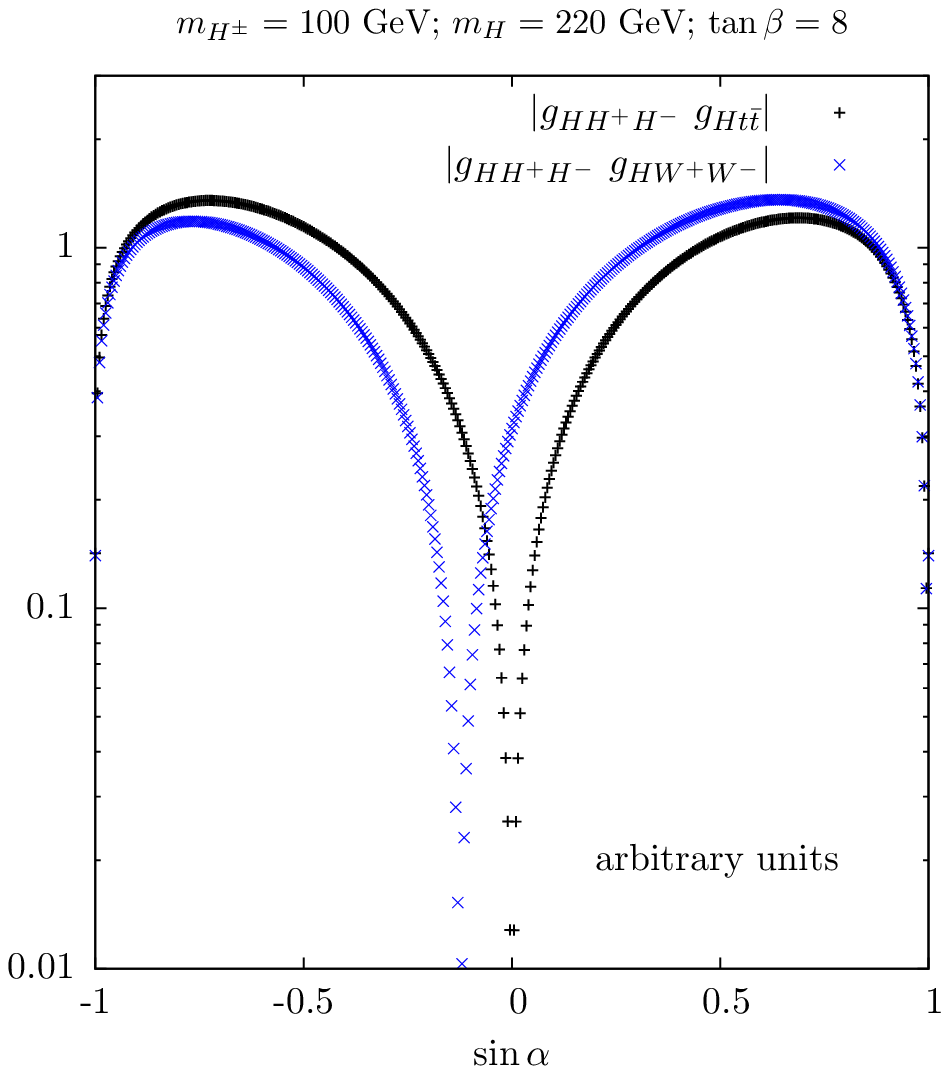}
\caption {Comparison between the product of the couplings $|g_{H H^+ H^-} \, g_{H t \bar t} |$, the main contribution in the gluon fusion process, 
and $|g_{H H^+ H^-} \, g_{H W^+ W^-} |$, the main contribution in the VBF process, for the values of the masses presented in the plot and 
$\tan \beta =3$ (left) and $\tan \beta =8$ (right). We have chosen $M=170$ $GeV$ and the values of $m_A$ and $m_h$ are irrelevant because the vertices do not depend on those masses.}
\label{fig:comp}
\end{figure}

\noindent
In figure \ref{fig:wwh} we present the total cross section for $pp \to H^+ H^-jj$ via VBF as a function of the charged Higgs boson mass. For a given 
$\tan \beta$, the values of $\sin \alpha$ were chosen so as to maximise the cross section while complying with all the constraints. As we saw before 
for the process $gg \to H^+ H^-$, here the lines also end abruptly due to the constraints imposed on the model. Also, like for $gg \to H^+ H^-$, 
the largest cross sections are obtained for large $\tan \beta$ which for $\tan \beta = 8$ are already of the order of the $pb$. In fact, it is clear 
from the structure of the couplings that VBF and gluon fusion have exactly the same behaviour with $\sin \alpha$ and $\tan \beta$ for large $\tan \beta$. 
The $HW^+ W^- $ coupling (appearing in VBF) is proportional to $\cos (\beta -\alpha)$ while the $H \bar t t$ one is proportional to $\sin \alpha / \sin 
\beta$ in the models under study (notice that for large $\tan \beta$,  $\cos (\beta -\alpha) \approx \sin \alpha / \sin \beta$). This behaviour is shown 
in figure \ref{fig:comp} where we compare the product of the couplings $|g_{H H^+ H^-} \, g_{H t \bar t} |$, the main contribution in the gluon fusion 
process, and $|g_{H H^+ H^-} \, g_{H W^+ W^-} |$, the main contribution in the VBF process (no constraints were applied). It is clear that, as 
$\tan \beta$ grows, the two processes become indistinguishable from the point of view of the 2HDM parameter space. Hence, if there is complementarity 
between the two processes this is only true for small $\tan \beta$. If a light charged Higgs boson is not found, then the small values of $\tan \beta$ 
will be excluded by $pp \to t \bar{t} \to  b \bar{b} W^\pm H^\pm$ as discussed in section \ref{sec:ttbar}. In this case, the region of the parameter 
space probed by VBF and gluon fusion will be almost the same.

\vskip -0.cm


\subsection{Associated production with a $W^\pm$}

\begin{figure}[h!]
\centering
\includegraphics[height=2.6in]{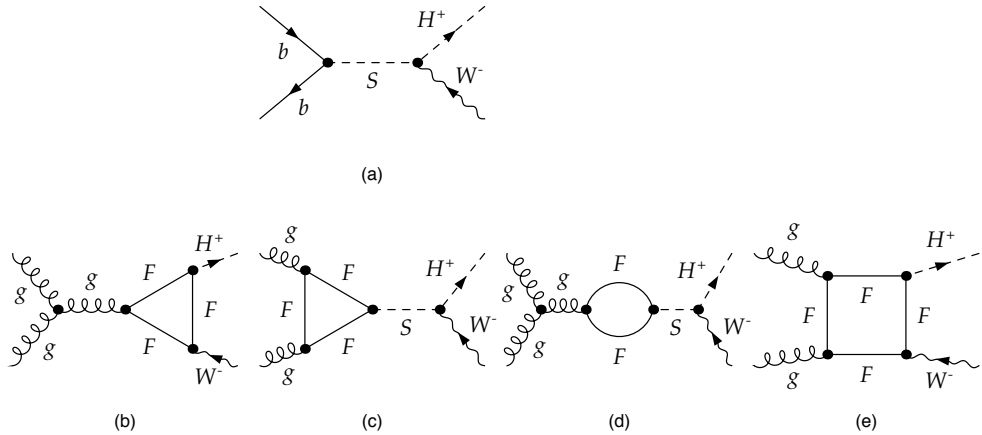}
\caption {Representative Feynman diagrams for $W^\pm H^\mp $ production via $ b \bar b$, (a), and $gg$, (b)-(e), fusion.
(Herein, $F$ represents a fermion, as appropriate, whereas $S$ a neutral Higgs boson,
as appropriate.)}
\label{fig:lproc2d}
\end{figure}
\vskip -0.0cm

\noindent
Contrary to charged Higgs boson pair production, discussed in the two previous subsections, the process $pp \to H^+ W^- + H^- W^+$, as shown in figure~\ref{fig:lproc2d}, 
does not depend on the Higgs self-couplings. Therefore, it is not suitable to explore the high $\tan \beta$ regime. The reason is simple: gauge couplings,
for large $\tan \beta$, are either proportional to $\sin \alpha$ or $\cos \alpha$. As an example, in the limit of very high $\tan \beta$, we have 
$|g_{H W^+ W-}| \approx | \sin \alpha \, \, g_{H W^+ W-}^{SM}|$ and thus the 2HDM coupling is always smaller than the corresponding SM coupling. However, 
as it was shown in~\cite{WHb} for the 2HDM, it is possible to have large values of the cross section for very large values of the charged Higgs boson 
mass. Hence, this process is particularly interesting in order to investigate the mass region near the threshold (notice once more we are considering only 
light charged Higgs bosons, that is, with masses below the $tb$ threshold).  

\begin{figure}[h]
\centering
\includegraphics[height=3.15in]{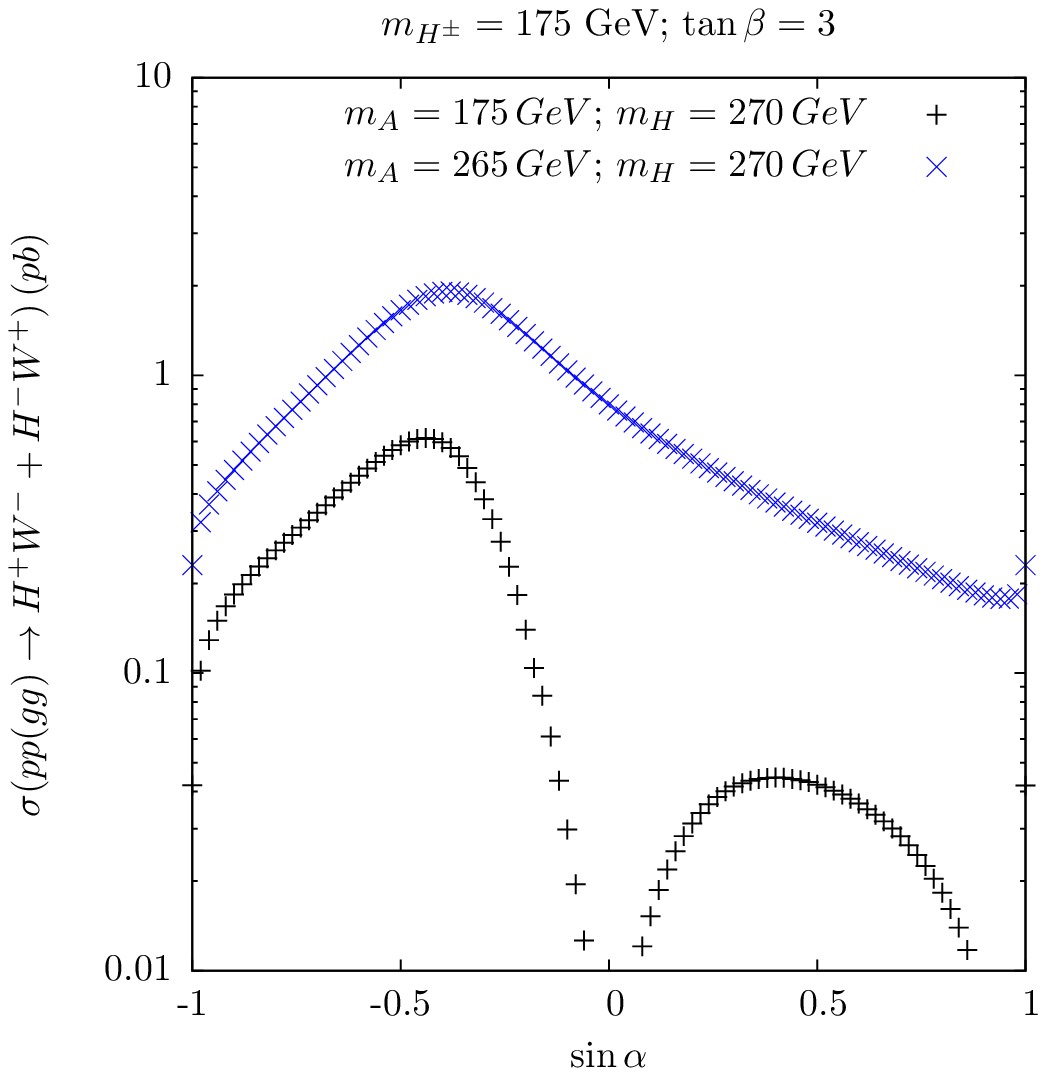}
\includegraphics[height=3.15in]{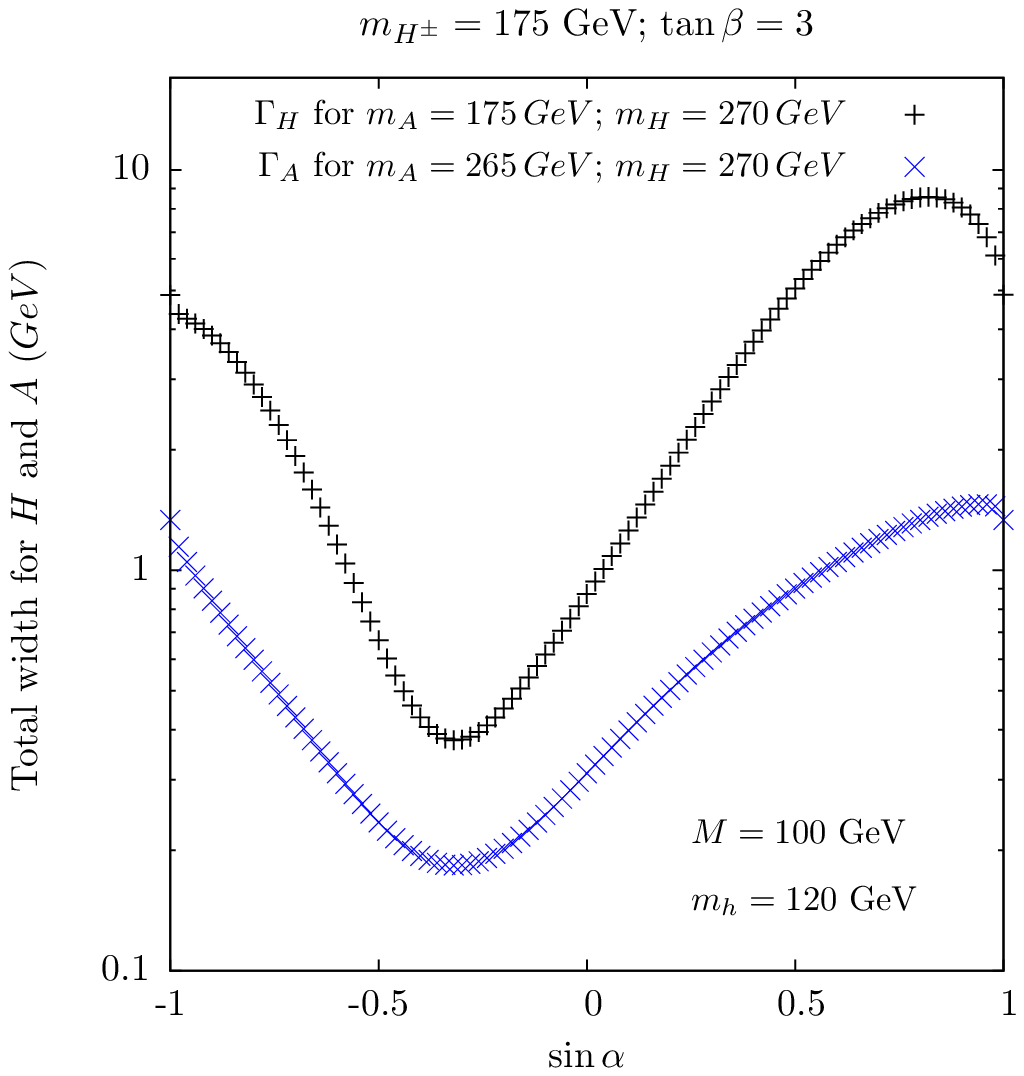}
\caption {In the left panel we show the production cross section for $pp (gg) \to H^+ W^- + H^- W^+ $ as a function of $\sin \alpha$ for $m_{H^\pm} =
175$ $GeV$ and $\tan \beta = 3$ (the remaining set of parameters is shown in the figures). In the right panel we present the total width of the heaviest CP 
even Higgs boson of the CP-odd Higgs boson for the same values of the model parameters.}
\label{fig:ggHW}
\end{figure}
\vskip -0.cm

\noindent
The most recent parton level analysis for $pp (gg+b\bar b)\to W^+ H^- +  H^+ W^-$ was performed in~\cite{WHc}. They have analysed the process $pp(b \bar b) \to W^+ H^- $, 
which has a phase space similar to the gluon initiated process when the triangle diagram dominates, followed by $H^\pm \to \tau \nu$. They have 
considered all possible decays of the $W^\pm$ boson. As compared to previous studies, they have now a more complete set of backgrounds that include 
$t \bar t$, $W^+W^-$ and $W^\pm +~jets$. Combining the analysis for the leptonic $W^\pm$ decay and hadronic $W^\pm$ decay cases, we conclude that a cross 
section of 1 $pb$ can be probed with approximately 6 $fb^{-1}$ of integrated luminosity and $\sqrt{s}=14$ $TeV$, for a charged Higgs mass of 175 $GeV$. 
We note however that, even with such a large cross section, the value of $S/B$ is close to 5\%. Nonetheless, the study presented in~\cite{WHc} clearly 
motivates a full simulation at detector level by the ATLAS and the CMS collaborations which would help to probe the region of charged Higgs masses near the threshold in these models.

\noindent
In the left panel of figure~\ref{fig:ggHW} we show the production cross section for $pp (gg) \to H^+ W^- + H^- W^+ $ as a function of $\sin \alpha$ 
for $m_{H^\pm} =175~GeV$ and $\tan \beta = 3$ (the remaining set of parameters is shown in the figures). In the right panel we present the total
 width of
the heaviest CP-even Higgs boson for the same values of the model parameters. First we note that, again, we have reached 
values of the cross sections of the level of the $pb$ with all constraints taken into account. The numbers become smaller if we move to larger values 
of $\tan \beta$. The values of the total width in the right panel show us how sensitive the cross section is to the width of the CP-even Higgs 
state, $H$, and 
especially to that of the CP-odd one,  $A$. As discussed for charged Higgs boson pair production, the values of the cross sections are 
extremely sensitive to the width of the resonant particle. This process has this interesting feature of involving the CP-odd Higgs state $A$ and the fact that 
in some regions of the parameter space the resonant $H$ and the resonant $A$ contributions are easily distinguishable. In fact,
because the CP-odd state
is not allowed to decay to a gauge boson pair, its width is always much smaller than that of a scalar state with same mass, thus making resonant production 
via an $A$ state the largest one. Taking the heaviest CP-even as an example, both couplings $g_{HW^\pm W^\mp}$ and $g_{H W^\pm H^\mp}$ are proportional 
to $\cos (\beta - \alpha)$ and therefore it is not possible to make one of them large and the other small simultaneously. In contrast, in the 
CP-odd case, the decay $A \to W^\pm H^\mp $ is usually the largest~\cite{Kanemura:2009mk} in most of the parameter space as long as it is 
kinematically allowed.

\section{Conclusions}

\noindent
In this work we have analysed all possible search modes for a light charged Higgs boson from a 2HDM at the LHC. We have started with processes that 
depend only on the charged Higgs boson mass and on $\tan \beta$. Using the ATLAS and CMS studies for $pp \to t \bar{t} \to  b \bar{b} W^\pm H^\mp$ we 
have drawn 95\% CL exclusion plots in the $(\tan \beta, m_{H^\pm})$ plane. For 30 $fb^{-1}$ of collected luminosity, the exclusion range spans from 
$\tan \beta \lesssim11$ for $m_{H^\pm}=90$ $ GeV$ to $\tan \beta \lesssim 6$ for $m_{H^\pm}=150$ $ GeV$ in model Type X and $\tan \beta \lesssim9$ for 
$m_{H^\pm}=90$ $ GeV$ to $\tan \beta \lesssim 2$ for $m_{H^\pm}=150$ $ GeV$ in model Type I. By performing a parton level analysis we have showed that the 
single top process deserves a full detector level analysis.  In fact, taking model Type X as an example, our parton level analysis shows that for a 
mass of 100 $ GeV$ an exclusion of  $\tan \beta \lesssim 5$ could be expected at 95\% CL for a collected luminosity of $10 \, fb^{-1}$ and $\tan \beta 
\lesssim 7$ for $30 \, fb^{-1}$ for model Type X. Although optimistic, the results show that a full detector level analysis would be worth performing.  
Combining the single top analysis with the one already performed for $t \bar t $, we expect to increase the excluded region in the $(\tan \beta, 
m_{H^\pm})$ plane. Finally, the last process that could contribute to improve the exclusion limits in the $(\tan \beta, m_{H^\pm})$ plane is direct 
charged Higgs boson production. We have shown that the cross sections are unfortunately too small and fall too rapidly with $\tan \beta$. Hence, no contribution is expected 
to help improving the above results. In table~\ref{tab:bench1} we present some benchmarks for the three processes that depend only on $\tan \beta$ and 
the charged Higgs boson mass. 

\begin{table}[h!]
\begin{center}
\begin{tabular}{c c c c c c c c c c c c c c c c c c c c c c c c c c c c} \hline \hline
$m_{H^{\pm}}$ ($GeV$) &    &&  100    &&   &&&&    &&  150    &&  \\ \hline 
$\tan \beta$              & 3 &&    10      && 30   &&&& 3 &&    10      && 30  \\ \hline \hline
$pp \to t \bar{t} \to  b \bar{b} W^\pm H^\mp$ \qquad \qquad  & Yes  && Yes  && No  &&&& Yes  && No  && No\\ \hline
$pp\rightarrow H^\pm b j$ \qquad \qquad  & Yes   && HL   && No   &&&& Yes   && No   && No \\ \hline
$pp (cs)  \to H^\pm (+j) $ \qquad \qquad  & No  && No   && No  &&&& No  && No   && No  \\ \hline \hline
\end{tabular}
\caption{Benchmarks for the three processes that depend only on $\tan \beta$ and on the charged Higgs mass. HL stands for High Luminosity.}
\label{tab:bench1}
\end{center}
\vskip -0.2cm
\end{table}

\noindent
If a charged Higgs boson is not found at the LHC, small values of $\tan \beta$ will be excluded with the above processes, as they rely only on the Yukawa 
couplings. Accessing the high $\tan \beta$ region is not possible regardless of the remaining 2HDM parameters. There are however some regions of the 
high $\tan \beta$ domain that can be probed at the LHC. We have shown that charged Higgs boson pair production via gluon fusion as well as 
VBF can give some scope but only in the scenarios of resonant production together with an enhancement of the Higgs self-couplings, namely the coupling 
between charged and neutral Higgs states. Parton level studies related to these processes lead us to the conclusion that the regions of large $\sin \alpha$ and 
large $\tan \beta$ give rise to cross sections of the order of the $pb$ that can be probed with just a few $fb^{-1}$ of integrated luminosity.  In table~\ref{tab:bench2}
 we present some benchmarks for the three processes where resonant production is allowed.

\begin{table}[h!]
\begin{center}
\begin{tabular}{c c c c c c c c c c c c c c c c c c c c c c c c c c c c} \hline \hline
$m_{H^{\pm}}$ ($GeV$) &    &&  100    &&   &&&&    &&  150    &&  \\ \hline 
$\tan \beta$              & 3 &&    10      && 30   &&&& 3 &&    10      && 30  \\ \hline \hline
$gg,b\bar b,q\bar q   \to H^+H^-$ \qquad \qquad  & Yes*  && Yes*  && Yes*  &&&& Yes*  && Yes*  && Yes*\\ \hline
$qQ        \to q'Q' H^+H^-$ \qquad \qquad  & Yes*   && Yes*   && Yes*   &&&& Yes*   && Yes*   && Yes* \\ \hline
$gg,b\bar b   \to H^\pm W^\mp$ \qquad \qquad  & Yes*  && HL*   && No  &&&& Yes*  && HL *  && No  \\ \hline \hline
\end{tabular}
\caption{Resonant production with enhancement of couplings. HL stands for High Luminosity.  (*In definite regions of the parameter space where resonant production is allowed.)}
\label{tab:bench2}
\end{center}
\vskip -0.2cm
\end{table}

\noindent
In conclusion, if one dismisses the usual presumption that a 2HDM can only be motivated within Supersymmetry,
thereby shifting the attention from its Type II realisation to other types, specifically to this work to the case
of Types I and X, one would find interesting phenomenology emerging at the current
LHC, manifesting itself in production and/or decay modes of light charged Higgs boson states,
i.e., below the top mass, that are possible neither in a Type II nor in a Type Y (the latter also known as III)
scenario. Very little luminosity may be necessary to ascertain the presence of such states at the CERN proton-proton
accelerator running at design energy (14 $TeV$) in a variety of novel signatures. In this paper, we have laid
the basis for a systematic exploration of such 2HDM types in the quest for the ultimate understanding
of the mechanism of EWSB. Sophisticated experimental analyses are now needed in order to finally confirm
or disprove the validity of these 2HDM hypotheses. To this end, we have produced computational tools, selection
procedures and benchmark scenarios that can readily be exploited in the LHC environment.  

\noindent {\bf Acknowledgements}
We thank Nuno Castro and  Filipe Veloso for discussions. We thank Oliver Brein for checking cross sections of gluon fusion processes.
All authors acknowledge financial support from the PMI2 Connect Initiative in the form of a Research Cooperation Grant. SM is 
partially supported by the NExT Institute. RG is supported by a Funda\c{c}\~ao para a Ci\^encia e Tecnologia Grant SFRH/BPD/47348/2008. 
MA is supported in part by Grant-in-Aid for Young Scientists (B) no. 22740137. SK is supported in part by Grant-in-Aid for Scientific Research (A) 
no. 22244031 and (C) no. 19540277. KY is supported by the Japanese Society for the Promotion of Science (JSPS Fellow (DC2)). RS is partially 
supported by an FP7 Reintegration Grant, number PERG08-GA-2010-277025. RG and RS are partially supported by PEst-OE/FIS/UI0618/2011.


\appendix
\section{Parton level analysis of charged Higgs production via single top production}

\noindent
In this Alppendix we present the highlights of the parton level analysis for the single top case. As mentioned before  we take as our signal the process
\begin{equation}
pp\rightarrow H^\pm b j\rightarrow b j\tau \nu_\tau.
\end{equation}
Regarding the background processes we consider the irreducible background processes
\begin{equation}
pp\rightarrow b j l(\tau,\mu,e) \nu_{\tau,\mu,e}
\end{equation}
and the reducible one
\begin{equation}
pp\rightarrow t\bar t\rightarrow W^+ b W^-\bar b
\end{equation}
where both $W^\pm$'s can decay semi-leptonic, fully leptonic or fully hadronic. 

\begin{figure}[h!]
\centering
\includegraphics[height=3.1in]{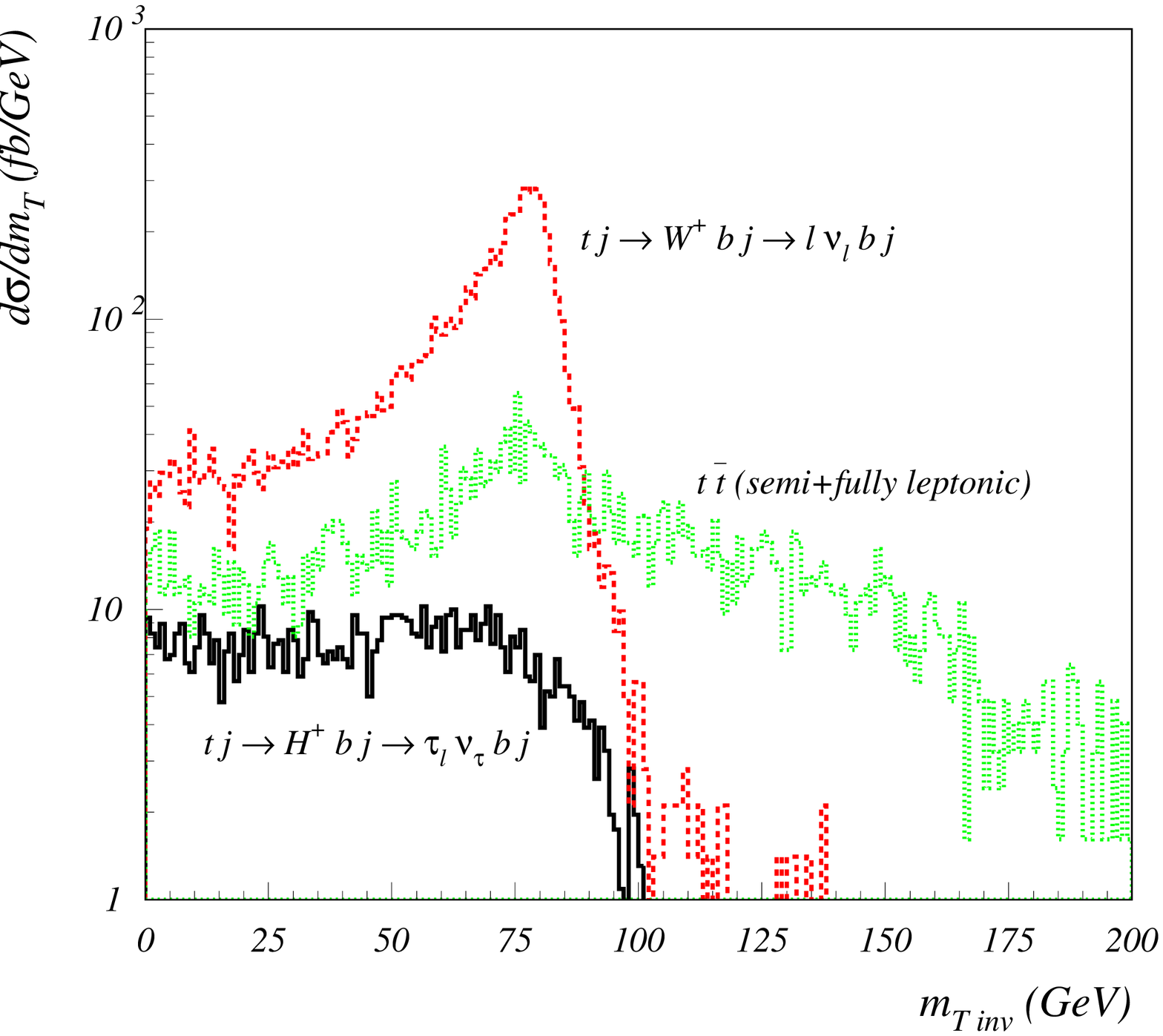}
\includegraphics[height=3.1in]{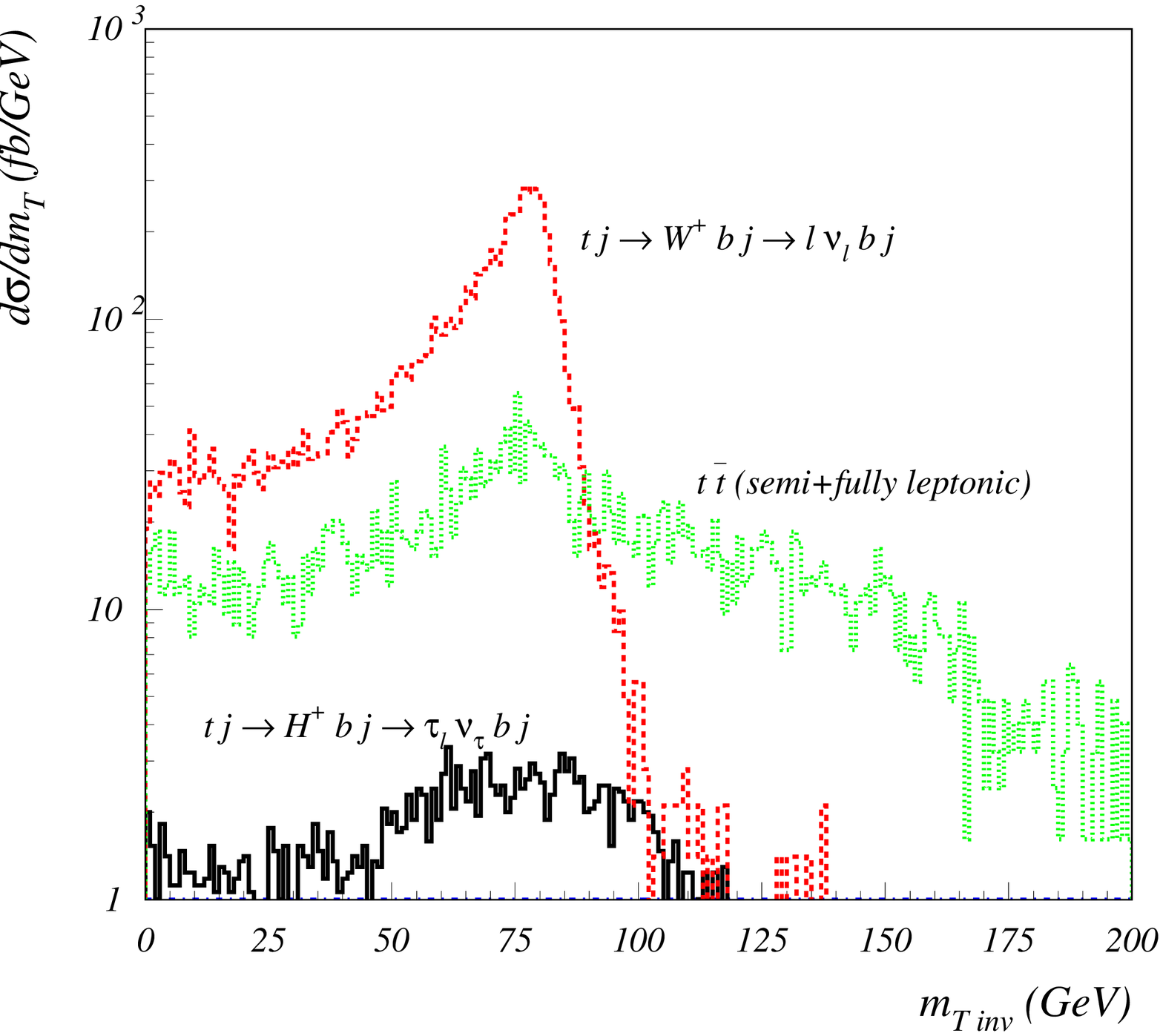}
\caption {In the left panel we show the transverse mass distribution for $m_{H^\pm}=100$  $ GeV$ and  $\tan\beta=1.5$ 
for signal and backgrounds. In the right panel we present the the same plot but for a charged Higgs boson mass of $m_{H^\pm}=140$ $GeV$.}
\label{fig:partonsingletop}
\end{figure}
\vskip -0.1cm

\noindent
First we have to eliminate the huge reducible background coming from $pp \to t \bar t$. We will consider processes with at least one lepton which 
means that we exclude the fully hadronic $t\bar t$ production background. The fully hadronic background is almost completely rejected by asking for a 
lepton with transverse momentum $p_T$ greater than $30$ $ GeV$ in the central detector region of $|\eta|\leq2.5$. The fully leptonic background is 
rejected by applying a veto on events that contain a second lepton with transverse momentum above $10$ $GeV$. These are the strategies developed in 
\cite{clemens} for SM single top production, which shows that the fully hadronic background can easily be reduced but should nevertheless be taken 
into account in a full detector level analysis. The semi-leptonic contribution is the hardest reducible background to deal with. In order to reduce 
it we apply a veto on events with more than two jets with transverse momentum greater than 15 $GeV$. This reduces the background to levels that are 
below the ones presented in~\cite{clemens} because we have no jets coming from hadronisation as this is a parton level analysis. Therefore, we have 
estimated in our semileptonic  $t\bar t$  background a reduction of a factor of 2 as compared to Ref. \cite{clemens} where, due the hadronisation of the 
jets, a veto is applied on events with more than four jets with transverse momentum greater than 15 $GeV$. Therefore our estimate of the total reducible 
background is probably optimistic by a factor of approximately 1.4. 

\begin{figure}[h!]
\centering
\includegraphics[height=3.5in]{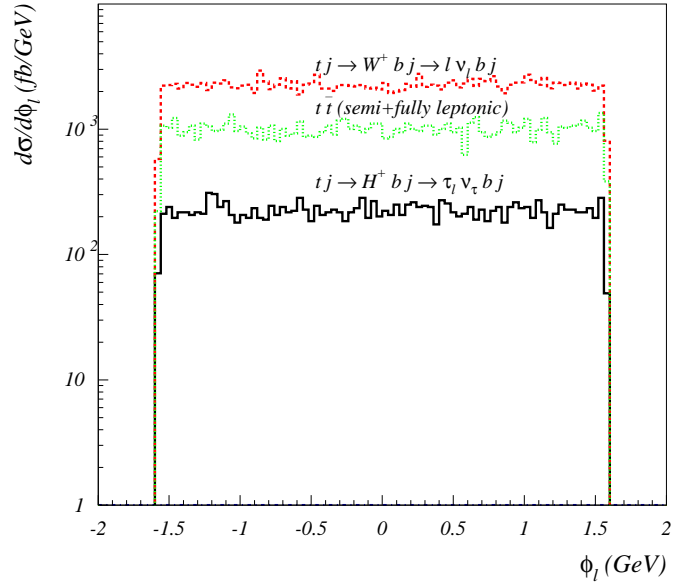}
\caption {Azimuthal angle distribution for $m_{H^\pm}=100$  $ GeV$ and  $\tan\beta=1.5$ for signal and backgrounds.}
\label{fig:partonsingletopb}
\end{figure}
\vskip -0.1cm

\noindent
Regarding the irreducible single top background we have looked at several distributions in order to find suitable cuts to minimise its effects. 
As expected, all $p_T$ distributions (of the lepton or any of the jets in the process) are very similar for both signal and irreducible background. 
The same is true for the missing energy distribution. Our hopes relied mainly on the lepton angular distributions (because of the different chirality 
in the couplings of the $W^\pm$ and $H^\pm$ bosons with the leptons) and on the transverse mass distribution as defined in~\cite{tmass}
\begin{equation}
M_T(l\nu)=\sqrt{2 |p_T^l| |p_T^{\rm{miss}}|-2 \vec p_T^{\, l}  . \vec p_T^{\, \rm{\, miss}}} \qquad 
\end{equation} 
where the superscript $l$ refers to electron and/or muon and the superscript "$\rm{miss}$" refers to total missing energy and total missing transverse momentum.
 In the left (right) panel of figure~\ref{fig:partonsingletop} we present the transverse mass distribution for $m_{H^\pm}=100 \, (140)$  $ GeV$ and  $\tan\beta=1.5$ 
for signal and backgrounds. It is clear that we can reduce drastically the background by avoiding the large transverse mass region. Therefore, we  
apply the following cut on the transverse mass, $m_T < 50$ $ GeV$, which optimises the significance for a 100 $ GeV$ Higgs boson. As the mass of the charged 
Higgs boson grows, this cut becomes less efficient because the signal maximum values move towards the $W^\pm$ mass peak as can be clearly seen in the 
right panel of figure~\ref{fig:partonsingletop}. Finally, all angular distributions related to the lepton have shown to be of no use in improving the 
analysis. As an example, we plot in figure~\ref{fig:partonsingletopb} the azimuthal angle distribution of the lepton (for $m_{H^\pm}=100$  $ GeV$ and  
$\tan\beta=1.5$). We believe that, because the leptons are highly boosted, the potential differences in the lepton angular distributions are washed out.
Conversely, the presence of undetectable particles in the final state prevents one from reconstructing the reference frame of the decaying boson.

\noindent
In summary, we have used the following set of cuts:
\begin{enumerate}
\item $\Delta R_{jj} > 0.4$ and $\Delta R_{jl} > 0.2$ where $j$ stands both for light jets and $b$-jets and $l$ is a lepton (electron or muon); 
\item we demand a lepton with transverse momentum greater than $p_T>30$ $ GeV$ in the central detector region of $|\eta|\leq2.5$;
\item to reject dilepton events a veto is applied on events that contain a second lepton with transverse momentum above $10$ $GeV$;
\item to reduce the QCD background, we apply a cut on the missing transverse momentum: we demand  $p_T^{\rm{miss}}>20$ $GeV$;
\item we demand events with at least two jets to have transverse momentum greater than $30$ $ GeV$ and to be in the region of the detector  with  $|\eta| \leq 4.5$;
\item a veto is applied  on events with more than two jets with transverse momentum greater than $15$ $GeV$;
\item we apply the following cut on the transverse mass: $m_T < 50$ $GeV$;
\item the efficiency to identify a lepton was chosen as  50\%;
\end{enumerate}
which led to the significances and $S/B$ ratios presented in table~\ref{tab:singtop}. We end this Appendix with a final comment on another important 
background, $W^\pm~+~n~jets$, which was shown by~\cite{clemens} to be negligible when the above set of cuts are imposed. The $W^\pm~+~n~jets$ background has a 
cross section two orders of magnitude larger than the $t \bar t$ one. If the $W^\pm$ decays hadronically this is just like all other QCD backgrounds  
and we can use a lepton and the missing transverse energy to discriminate against events containing only jets. If the $W^\pm$ boson decays leptonically,  
we can reject about 95\% of all remaining $W^\pm~+~n~jets$ events by demanding a $b$-tagged jet (which we do by demanding a $b$-jet with a transverse  
energy larger than 30 $GeV$). According to~\cite{clemens},  the $W^\pm~+~jets$ background should be about 10\% of the total background in the case of 
SM single top production.

\section{Charged Higgs boson pair production in left-right symmetric models}

\noindent
A study of charged Higgs pair production in the context of a left-right symmetric model was first performed in~\cite{Datta:1999nc}.
A particular final state, $e^\pm \mu^\mp + \slashed{p}_T$, was chosen and a significance of $\sigma_S/\sqrt{\sigma_B} \approx 1.2 \, fb^{1/2}$ was obtained, for a charged Higgs boson mass of 100 $GeV$. 
The large value for the signal cross section is mainly due to the values of the BRs - in these models the charged Higgs boson decays to a lepton and an anti-neutrino 
independently of the lepton family.  More recently, see~\cite{Davidson:2010sf}, a similar study was performed in the context of a 2HDM-like model, where three 
gauge-singlet right-handed Weyl spinors were added to become the right-handed components of the three Dirac neutrinos. The authors
of this paper have investigated three 
possible final states, $e^+ e^- + \slashed{p}_T$, $\mu^+ \mu^- + \slashed{p}_T$ and $e^\pm e^\mp + \slashed{p}_T$, for charged Higgs 
boson masses of 100 and 300 $ GeV$ and in the scenario where BR$(H^+ \to e^+ \nu)$ = BR$(H^+ \to \mu^+ \nu) = 1/3$.  Combining all analyses 
for this particular scenario they conclude that a 20 $fb^{-1}$ luminosity is needed for a discovery with a $5 \sigma$ significance for a 100 $ GeV$ 
charged Higgs boson. We will now use their results to make an estimate of the cross sections that can be probed in both models Type I and X. 
Before proceeding we should however note that these are truly crude estimates, i.e., although the final states are the same and the production process looks 
similar, there are differences due to the structure of the couplings, e.g.,  it is clear that the amount of missing energy present, that would reflect 
in the signal efficiency, is different in each case. Nevertheless we believe we can use their results as a guide until a parton level analysis for 
this specific process is available. Finally, before presenting their results applied to our models,  we should mention that the final states chosen 
in~\cite{Datta:1999nc,Davidson:2010sf} are not the ones that would give the largest number of signal events.
\begin{table}[h!]
\begin{center}
\begin{tabular}{c c c c c c c c c c c c} \hline \hline
 $\tan \beta$ $(m_{H^{\pm}} = 100 \, GeV)$  & 1 &&  3 && 30   \\ \hline \hline
${\rm{BR}}_I (H^{\pm} \to \tau_l \nu) \times {\rm{BR}}_I (H^{\pm} \to cs)$ \qquad \qquad  & 8\%  && 8\%  && 8\%  \\ \hline
${\rm{BR}}_I (H^{\pm} \to \tau_l \nu) \times {\rm{BR}}_I (H^{\pm} \to \tau \nu)$ \qquad \qquad  & 28\%   && 28\%   && 28\%   \\ \hline
${\rm{BR}}_X (H^{\pm} \to \tau_l \nu) \times {\rm{BR}}_X (H^{\pm} \to cs)$ \qquad \qquad  & 8\%  && 0.2\%   && $\approx$ 0  \\ \hline
${\rm{BR}}_X (H^{\pm} \to \tau_l \nu) \times {\rm{BR}}_X (H^{\pm} \to \tau \nu)$ \qquad \qquad & 28\%   && 58\%  && 58\%  \\ \hline \hline
\end{tabular}
\caption{Product of $\tau$ BRs for a charged Higgs boson mass of 100 $ GeV$ in models Type I and X and for three values of $\tan \beta$. The largest 
numbers occur for the case when one $\tau$ decays leptonically and the other hadronically (we are considering that we need at least one $\tau$ do decay 
leptonically for triggering purposes).}
\label{tab:brstaus}
\end{center}
\vskip -0.5cm
\end{table}
In fact, taking into account the values for the BRs presented in table~\ref{tab:brstaus}, a final state where 
one $\tau$ decays leptonically and the 
other hadronically is obviously more appropriate to study models Type I and X.

\noindent
In 2HDMs Type I and X, and for $\tan \beta \gtrsim 2$, the main decay of the charged Higgs boson is $H^+ \to \tau^+ \nu$ and decays to 
$\mu \nu$ and $e \nu$ are negligible. Then, using the scenario in~\cite{Datta:1999nc,Davidson:2010sf}, both $\tau$'s have to decay leptonically to 
an electron or a muon plus missing energy with SM rates. The number of signal events is given by
\begin{eqnarray}
\sigma (pp \to H^+ H^-) \, {\rm{BR}} (H^+ \to \tau^+ \nu) \, {\rm{BR}} (H^- \to \tau^- \nu)  \, {\rm{BR}} (\tau^+ \to l^+ \nu) \, {\rm{BR}} (\tau^- \to l'^- \nu)
\end{eqnarray}
where $l,l'=\mu,e$. The BR$(\tau^+ \to l^+ \nu)$ is $\approx 17\%$ for muons and $\approx 18\%$ for electrons. For $\tan \beta \gtrsim 2$, 
${\rm{BR}} (H^+ \to \tau^+ \nu) $ is $69\%$ in model Type I and close to $100\%$ in model Type X. Taking the previous studies~\cite{Datta:1999nc,Davidson:2010sf} as 
a guide, we would conclude that a significant cross section is one of the order of 400 $fb$. This means that, using their analysis for our signal 
despite the difference in, e.g., the total missing energy (note that the background is the same), one would be able to start probing a significant 
parameter region with 30 $fb^{-1}$ for 2HDM Type X if the production cross section was of the order of 400 $fb$.


\end{document}